\documentclass[aps,prb,floatfix,twocolumn,showpacs,longbibliography]{revtex4-2}
\usepackage{amsmath}
\usepackage{url}
\usepackage[ruled,vlined]{algorithm2e}
\usepackage{algorithmic}
\usepackage{graphicx}% Include figure files
\usepackage{dcolumn}% Align table columns on decimal point
\usepackage{bm}% Bold math
\usepackage{amssymb}% Bold math
\usepackage{rotating}% Rotate figures and tables
\usepackage[abs]{overpic}% Over pic axis
\usepackage{xcolor}% Over pic axis color
\usepackage{tabularx}% Table adjustment
\usepackage[justification=RaggedRight]{caption}
\usepackage{hyperref}
\usepackage[utf8]{inputenc}

\usepackage{braket}
\usepackage{soul}
\usepackage[english]{babel}
\hypersetup{colorlinks = true, citebordercolor={blue}, linkcolor={blue}, citecolor={blue}, urlcolor={blue}}
\usepackage{subcaption}
\usepackage{overpic}
\newcommand{\tc}{\textcolor{black}}
\begin{document}
%\title{Geometric Analysis of the Dual Condensate Hamiltonian}
%\title{First- and Second-order Quantum Phase Transitions between Superconductivity and Exiton Condensation in a Model for their Coexistence}
%\title{Quantum Phase Transitions between Superconductivity and Exciton Condensation in a Hamitonian Model for their Entangled Coexistence}
%\title{Quantum Phase Transitions between Superconductivity and Exciton Condensation in a Hamiltonian Model for Simultaneous Exciton and Fermion-Pair Condensations}
%\title{Quantum Phase Transitions in a Model Hamiltonian Exhibiting Entangled Simultaneous Exciton and Fermion-Pair Condensations}
\title{Quantum Phase Transitions in a Model Hamiltonian Exhibiting Entangled Simultaneous Fermion-Pair and Exciton Condensations}
\author{Samuel Warren, LeeAnn M. Sager-Smith, and David A. Mazziotti}
\email{damazz@uchicago.edu}
\affiliation{Department of Chemistry and The James Franck Institute, The University of Chicago, Chicago, IL 60637}
\date{Submitted June 6, 2022\tc{; Revised August 13, 2022}}
\begin{abstract}
Quantum states of a novel Bose-Einstein condensate, in which both fermion-pair and exciton condensations are simultaneously present, have recently been realized theoretically in a model Hamiltonian system.  Here we identify quantum phase transitions in that model between fermion-pair and exciton condensations based on a geometric analysis of the convex set of ground-state 2-particle reduced density matrices (2-RDMs).  The 2-RDM set provides a finite representation of the infinite parameter space of Hamiltonians that readily reveals a fermion-pair condensate phase and two distinct exciton condensate phases, as well as the emergence of first- and second-order phase transitions as the particle number of the system is increased.  The set, furthermore, shows that the fermion-exciton condensate (FEC) lies along the second-order phase transition between the exciton and fermion-pair condensate phases.  The detailed information about the exciton and fermion-pair phases, the forces behind these phase, as well as their associated transitions provides additional insight into the formation of the FEC condensate, which we anticipate will prove useful in its experimental realization.
\end{abstract}
\maketitle
\section{Introduction}
Condensation phenomena were some of the first macroscopic-quantum effects observed in the modern physics era with the discovery of superconductivity in mercury \cite{onnes_investigations_1913} and helium-4's superfluidity\cite{kapitza_viscosity_1938} in the early 1900's. These exotic phases of matter continue to dominate research in fields concerned with the development of breakthrough technologies like high-temperature superconductors \cite{hartnoll_building_2008,anderson_twenty-five_2013,chen_atomic_2020,zhang_design_2022}, superconducting \cite{bouchiat_quantum_1998,nakamura_coherent_1999} and topological \cite{kitaev_fault-tolerant_2003, sau_generic_2010, dusuel_robustness_2011} qubits, and excitonic-superconducting materials \cite{keldysh_coherent_2017,fil_electron-hole_2018,schouten_exciton_2021,muraviev_exciton_2022}. Additionally, these efforts have lead to the creation of new theories regarding quantum phase transitions \cite{sachdev_quantum_2011} and novel model Hamiltonians in order to explore these exotic forms of matter on a simplified manifold.

Recent work by Sager and Mazziotti \cite{sager_simultaneous_2022} has demonstrated the existence of simultaneous exciton and fermion-fermion pair condensation or Fermion-Exciton Condensation (FEC) in a model Hamiltonian. This was achieved by combining two other model Hamiltonians that demonstrate condensation and long range off-diagonal order. The first, the Bardeen-Schrieffer-Cooper (BCS) Hamiltonian \cite{bardeen_theory_1957}, was created with the express purpose of modeling fermion-fermion pair condensation, earning them a Nobel prize in 1972. The second model, the Lipkin-Meshkov-Glick (LMG) Hamiltonian \cite{lipkin_validity_1965, meshkov_validity_1965, glick_validity_1965,mazziotti_exactness_2004}, was originally constructed with the goal of testing quantum-many body approximations on an exactly solvable, but highly correlated system, and has been extensively studied for its phase transitions \cite{heiss_large_2005,castanos_classical_2006,gidofalvi_computation_2006,chen_unconventional_2006, romera_phase_2014, wang_characterizing_2021}. For the purposes of the FEC Hamiltonian, it has been observed that many of the LMG ground-states have significant exciton condensation character \cite{sager_preparation_2020}. The model that results from combining these two systems generates states that demonstrated FEC. However, while it was demonstrated that these FEC states exist in the thermodynamic limit \cite{sager_potential_2020}, the underlying phase behavior that gives rise to these states remained obscure.

\tc{To elucidate the phases and their transitions in an FEC system, we draw upon two-electron reduced-density-matrix (2-RDM) techniques for quantum phase transitions~\cite{wu_quantum_2004,gidofalvi_computation_2006,schwerdtfeger_convex-set_2009,zauner_symmetry_2016} in combination with 2-RDM-based signatures for the condensations~\cite{yang_concept_1962, sasaki_eigenvalues_1965, coleman_structure_1965, garrod_particlehole_1969, safaei_quantum_2018}.}  As demonstrated by Erdahl and Jin \cite{erdahl_lower_2000}, Gidofalvi and Mazziotti \cite{gidofalvi_computation_2006}, Zauner and Verstraete \cite{zauner_symmetry_2016}, and recently, the authors on a quantum computer~\cite{Warren2022}, a geometric analysis of the ground-state set of 2-RDMs provides a visualization of symmetry breaking and phase transitions in both classical and quantum systems, which hearkens back to the geometric approach developed by Gibbs and Maxwell by generalizing Maxwell eponymous surface to discrete systems. This method provides a generalizable geometric framework for quantum phase transitions in terms of the convex set of 2-RDMs that has two important advantages: (1) based on a quantum information perspective, the 2-RDM theory relies upon the state space of all two-body observables rather than a specific Hamiltonian to examine the transition, and (2) it reduces the analysis of an infinite space of Hamiltonians to the study of recognizable geometric features like planes or ruled surfaces in the finite and convex set of 2-RDMs. Such a three-dimensional analysis allows for visualizing a greater swath of the space of all possible Hamiltonians than traditional single-order parameter or energy-level analysis. As a result, this higher level prospective can guide focused studies with these more traditional techniques to regions with interesting critical behavior.

%Here, through exactly solving the system Hamiltonian for finite systems, we are able to determine various phases of exciton and fermion-fermion pair condensation, as well as identify the emergence of phase transitions with growing particle number. These phase transitions prove particularly interesting as the FEC states are found to lie in these critical regions, and they provide evidence that long range off-diagonal coupling, due to scattering, generates and can even suppress simultaneous condensation. This analysis supplies a perspective, which may aid in finding real-world systems that exhibit simultaneous condensation, and exploit their potential in energy transport and electronics.

\tc{In this paper we determine the quantum phases and their transitions in the FEC system described by the model Hamiltonian developed in Ref.~\cite{sager_simultaneous_2022}.  We identify three discrete phases of 2-body condensates---two fermion-pair condensates and an exciton condensate---, as well as the emergence of phase transitions between these regions with growing particle number. We characterize the order of these transitions and provide a map of the entire phase space using 2-RDM theory. These phase transitions prove particularly interesting as the FEC states are found to lie in the critical regions. However, due to certain system symmetries, we can prove, through a novel extrapolation of the ground-state wave functions to the thermodynamic limit, that one of the exciton condensate regions is fundamentally incapable of sustaining FEC states as it transitions to the fermion-pair condensate region. Additionally, this analysis supplies a perspective, which may aid in finding real-world systems that exhibit simultaneous condensation, and exploit their potential in energy transport and electronics.}

\section{Theory}

We cover the signatures or signs of condensation in Reduced Density Matrix theory in section~\ref{sec:T1}, then provide a description of the Fermion-Exciton Condensate Hamiltonian in section~\ref{sec:T2}, and finally, identify the geometric consequences on the convex set of ground-state 2-RDMs of quantum phase transitions in section~\ref{sec:T3}.

\subsection{RDM Signatures of Condensation} \label{sec:T1}
\tc{Bosonic condensation is the result of multiple bosonic particles occupying the same energy orbital. Fermions, in contrast, are limited by the Pauli Exclusion principle to only one particle per orbital. This difference can be readily detected by comparing the one-boson reduced density matrix to the one-fermion reduced density matrix (1-RDM)
\begin{equation}
    ^1D^i_j = \bra{\Psi}{\hat a}^\dagger_i{\hat a}_j\ket{\Psi}
\end{equation}
where ${\hat a}^\dagger_i$ and ${\hat a}_i$ are the fermionic creation and annihilation operators for the $i^{\rm th}$ orbital acting on the $N$-fermion wave function $\ket{\Psi}$, and the bosonic 1-RDM can be obtained by swapping these elements for their bosonic counterparts. The eigenvalues of these matrices indicate the occupation of a specific energy orbital, and therefore the fermionic eigenvalues are strictly less than or equal to 1, while the bosonic eigenvalues have no such restriction \cite{penrose_bose-einstein_1956}. }

Fermion-pair condensation---superconductivity being a prominent example---is the result of multiple quasi-bosonic fermion-fermion pairs occupying a single two-fermion function, known as a geminal, which is analogous to the single-fermion occupation of an orbital \cite{bardeen_theory_1957,bose_plancks_1924,surjan_introduction_1999}. Such condensation results in the frictionless flow, superfluidity, of these particle-particle pairs through the material \cite{yang_concept_1962,london_bose-einstein_1938,tisza_theory_1947}. In the case of Cooper (electron-electron) pairs \cite{bardeen_theory_1957}, superfluidity manifests as superconductivity, which has a vast set of potential applications including in energy transport \cite{anderson_twenty-five_2013} or in the nascent field of quantum computing \cite{bouchiat_quantum_1998,nakamura_coherent_1999}.

In order to verify the presence and extent of fermion-fermion pair condensation, Yang \cite{yang_concept_1962} and Sasaki \cite{sasaki_eigenvalues_1965} independently developed a computational signature derived from the particle-particle reduced density matrix (RDM), $^2D$, whose elements are described by
\begin{equation}
	^2D^{i,j}_{k,l} = \bra{\Psi}{\hat a}^\dagger_i{\hat a}^\dagger_j{\hat a}_l{\hat a}_k\ket{\Psi}.
\end{equation}
Eigenvalues of the $^2D$ matrix describe the occupation of the two-fermion geminals \cite{coleman_structure_1965,raeber_large_2015}, meaning that when one of these eigenvalues exceeds the Pauli-like limit of one, multiple fermion-fermion pairs occupy a single geminal and hence fermion-pair condensation occurs. More generally, measurement of the largest eigenvalue, $\lambda_D$, serves as a indicator of off-diagonal long-range order in a system \cite{raeber_large_2015}.

Similarly, exciton condensation occurs when multiple fermion-hole pairs begin to occupy the same particle-hole function, resulting in the superfluidity of the quasi-bosonic particle-hole pairs \cite{keldysh_coherent_2017,fil_electron-hole_2018}. The computational signature of this condensation is the second largest eigenvalue of the particle-hole RDM or the G matrix \cite{safaei_quantum_2018} where the largest eigenvalue is the ground-state-to-ground-state transition. This transition can be removed using the one-fermion RDM, $^1D$:
\begin{multline}
	^2\tilde{G}^{i,j}_{k,l}= ^2G^{i,j}_{k,l}-^1D^i_j \text{ }^1D^l_k\\
	=\bra{\Psi}{\hat a}^\dagger_i{\hat a}_j{\hat a}^\dagger_l{\hat a}_k\ket{\Psi}-\bra{\Psi}{\hat a}^\dagger_i{\hat a}_j\ket{\Psi}\bra{\Psi}{\hat a}^\dagger_l{\hat a}_k\ket{\Psi},
\end{multline}
leaving the largest eigenvalue of the modified G matrix, which we denote as $\lambda_G$, as a signature of exciton condensation.

These signatures have been successfully used to identify condensation in a variety of systems \cite{safaei_quantum_2018,sager_superconductivity_2021,sager_cooper-pair_2022,schouten_exciton_2021,sager_beginnings_2022}, and because of the linear mapping between $^2D$ and $^2G$ \cite{mazziotti_quantum_2006},
\tc{
\begin{equation}
    {}^2G^{i,j}_{k,l} = \delta^{l}_{j} {}^{1}D^{i}_{k} -{}^{2}D^{i,l}_{k,j}
\end{equation}
where $\delta$ is the Kronecker delta function,} calculation of one type of reduced density matrix trivializes the calculation of the other one. This relationship makes it possible to easily determine if both forms of condensation exist within a single system by calculating the 2-RDM---the $^2D$ matrix---either exactly or through approximate methods \cite{mazziotti_realization_2004,mazziotti_anti-hermitian_2006,mazziotti_large-scale_2011}.

\subsection{Fermion-Exciton Condensate Hamiltonian}

\label{sec:T2}

The Fermion-Exciton Condensate (FEC) model Hamiltonian was proposed by Sager and Mazziotti \cite{sager_simultaneous_2022} as a means to produce a model capable of demonstrating exciton condensation, fermion-pair condensation (FPC), and simultaneous FEC. The Hamiltonian is constructed by combining two model systems, the Bardeen-Schrieffer-Cooper (BCS) or Pair-Force (PF)\cite{bardeen_theory_1957} and Lipkin-Meshkov-Glick (LMG) \cite{lipkin_validity_1965,meshkov_validity_1965, glick_validity_1965} systems, which independently are able to achieve fermion-pair and exciton condensation, respectively. The resulting Hamiltonian is
\begin{equation}
	H = \epsilon\hat{E}+\frac{\lambda}{2} \hat{\Lambda}+ \frac{w}{2} \hat{W} - g \hat{G}
	\label{eq:ham}
\end{equation}
where
\begin{equation}
	\hat{E}=\sum_{p=1}^N \left({\hat a}^\dagger_i{\hat a}_i-{\hat a}^\dagger_{i+N}{\hat a}_{i+N}\right),
	\label{eq:epsilon}
\end{equation}
\begin{equation}
	\hat{\Lambda} = \sum_{q,p=1}^N\left( {\hat a}^\dagger_p{\hat a}^\dagger_q{\hat a}_{q+N}{\hat a}_{p+N} +{\hat a}^\dagger_{p+N}{\hat a}^\dagger_{q+N}{\hat a}_q{\hat a}_p\right), \\
	\label{eq:lambda}
\end{equation}
\begin{equation}
	\hat{W} = \sum_{q,p=1}^N\left( {\hat a}^\dagger_{p+N}{\hat a}^\dagger_q{\hat a}_{q+N}{\hat a}_{p}\right),\\
	\label{eq:W}
\end{equation}
and
\begin{equation}
	\hat{G} = \sum_{p,q=1}^N{\hat a}^\dagger_{2p-1}{\hat a}^\dagger_{2p}{\hat a}_{2q}{\hat a}_{2q-1} .
	\label{eq:G}
\end{equation}
This Hamiltonian describes a system of spinless fermions with two energy levels each of which contains $N$ orbitals. Within these levels, there is a pairing force between sets of adjacent $\left(2k-1, 2k\right)$ orbitals, $\hat{G}$, derived from the BCS Hamiltonian. This pair force can also move these pairs of particles to other sets of empty adjacent orbitals. Additionally, the LMG model introduces several terms, the first of which, $\hat{E}$, is set to zero in this study as it is unnecessary to maximize either form of condensation. \tc{In fact, because both condensates are two-body phenomena, the one-body $\hat{E}$ term counteracts the condensation by localizing particles in specific orbitals or decreasing the importance of off-diagonal coupling in the Hamiltonian.  Often quantum phase transitions arise from a competition between one- and two-body terms in the Hamiltonian, but in this case the competition is between two distinct two-body terms in the Hamiltonian that favor different types of pairing.} The $\hat{\Lambda}$ scattering force moves pairs of particles between the energy levels, and the $\hat{W}$ scattering force interchanges particles between the levels, leaving the occupation number within each level unchanged. The eigenvalues of the resulting ground-state 2-RDMs are bounded from above by $\lambda_D\leq \frac{N}{2}\left(1-\frac{N-2}{r}\right)$ \cite{coleman_structure_1963} and $\lambda_G\leq\frac{N}{2}$ \cite{garrod_particlehole_1969} where $r$ is the number of orbitals or $2N$ in this system.

\subsection{Geometry of the Set of 2-RDMs}

\label{sec:T3}

Exploration of the FEC system in this paper is done with methods recently developed within the field of quantum information theory that utilize the convexity of the set of ground-state 2-RDMs to provide a visually compact overview of the infinite space of Hamiltonians \cite{schwerdtfeger_convex-set_2009,zauner_symmetry_2016,gidofalvi_computation_2006,chen_geometry_2016}. Analyzing the geometry of the resulting structure can give significant information about quantum criticality in the system, even in the finite particle limit \cite{chen_geometry_2016}. Abstract concepts like symmetry breaking become readily visible in the form of ruled surfaces on the convex set, and discontinuities in the surface of the set serve as an indication of first-order phase transitions. Additionally, this approach allows for the identification of critical phenomenon by observing the `speed' of the RDM along the edge of the set moving between Hamiltonian configurations $H_i$ and $H_f$ linearly in the space of Hamiltonians as described by
\begin{equation}
	H_t=H_i\left(1-\chi\right)+\chi H_f
	\label{eq:traj}
\end{equation}
This `speed' or curvature of the set is then defined as $(v\cdot v)^{1/2}$ with
\begin{equation}
	v=\left(\frac{\partial \langle \hat{\Lambda}\rangle}{\partial \chi},\frac{\partial \langle \hat{W} \rangle}{\partial \chi},\frac{\partial \langle \hat{G}\rangle}{\partial \chi}\right).
	\label{eq:speed}
\end{equation}
\tc{where $\langle \hat{O}\rangle$ is the expectation value of the observable $\hat{O}$}. These methods can be used to provide a high-level overview of the system, by compressing the infinite space of Hamiltonians into a convex set that emphasizes regions of interest through easily discernible visual cues. This then allows for a more focused study of the areas of interest with traditional methods like energy-level analysis.

In order to characterize phase behavior as well as the exciton and fermion-pair condensation within the system, the ground-state RDMs for the finite-particle systems are solved \tc{by numerical diagonalization of the Hamiltonians.  For code to perform the construction of the Hamiltonians and the calculation of the ground-state RDMs refer to Ref.~\cite{dc}.}  The D and G matrices are then analyzed as discussed above to provide an overview of the critical phenomenon in the system.

\begin{figure*}
	\centering
	\begin{subfigure}{.49\linewidth}
%		\begin{overpic}[width = \linewidth]{4p3rdmD.png}
		\begin{overpic}[width = \linewidth, trim = {6.3cm 7cm 6cm 9cm},clip]{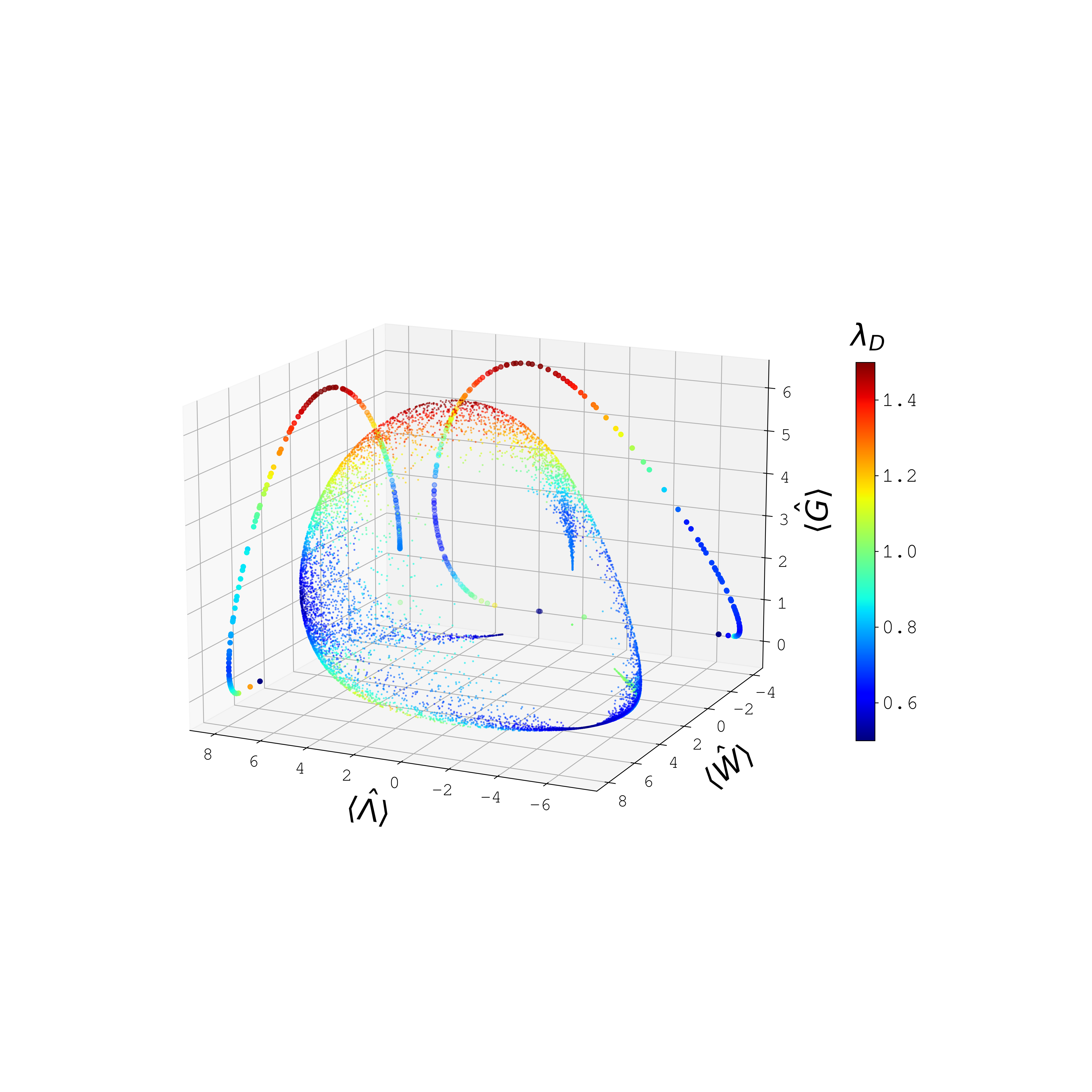}
%			\put(10,170){\normalsize(a)}
			\put(0,180){\normalsize(a)}
		\end{overpic}
		\phantomcaption
		\label{fig:lambdaD}
	\end{subfigure}
	\hfill
	\begin{subfigure}{.49\linewidth}
%		\begin{overpic}[width = \linewidth]{4p3rdmG.png}
		\begin{overpic}[width = \linewidth, trim = {6.3cm 7cm 6cm 9cm},clip]{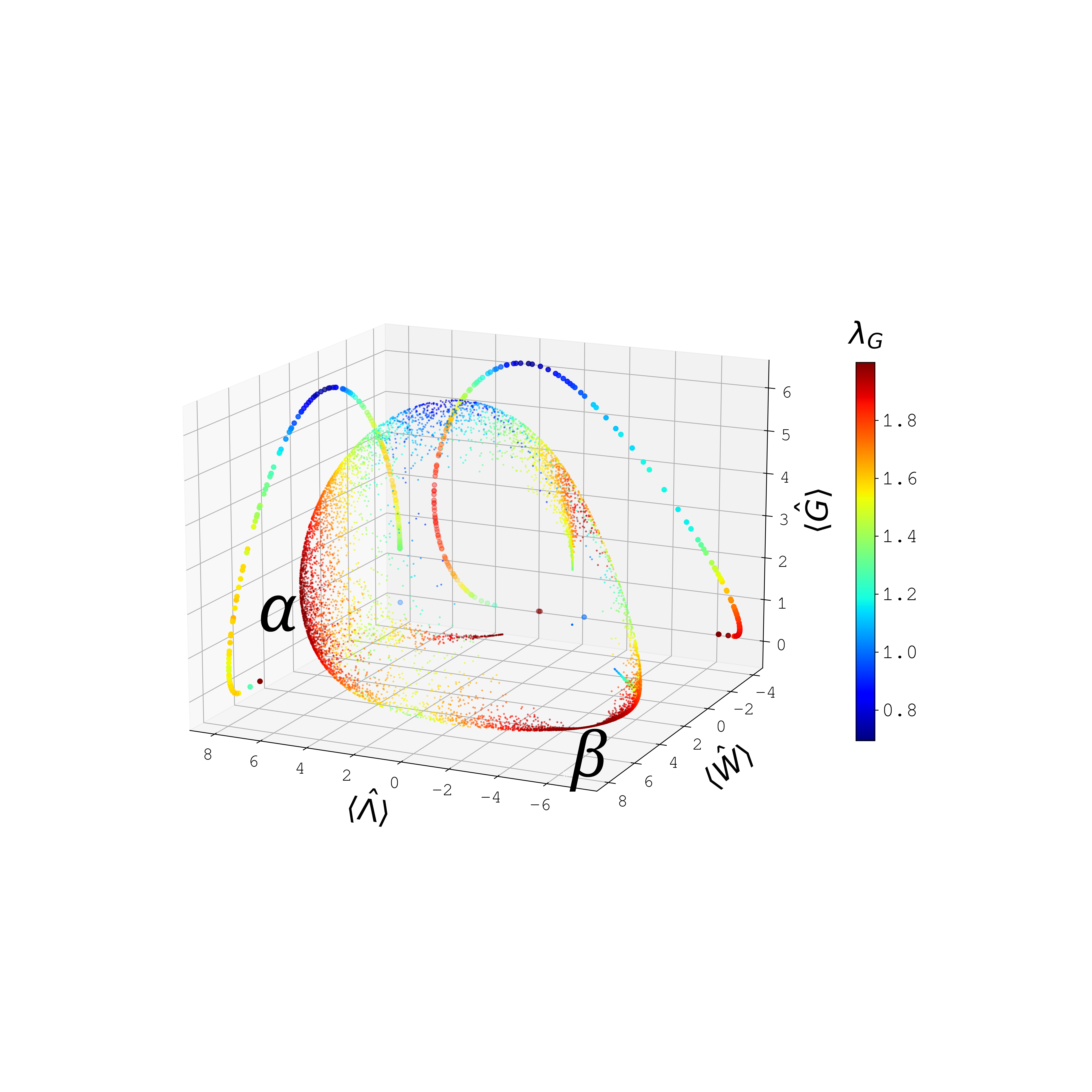}
%			\put(10,170){\normalsize(b)}
			\put(0,180){\normalsize(b)}

		\end{overpic}
		\phantomcaption
		\label{fig:lambdaG}
	\end{subfigure}
		\vspace{-1.1cm}
	\caption{Color map of Fermion-Pair and Exciton Condensation on the Convex Set of 2-RDMs. This scatter plot shows the resulting RDMs from a random sampling of Hamiltonian, Eq.\ref{eq:ham}, configurations \tc{for N=4 particles} with a color map indicating the value of $\lambda_G$ and $\lambda_D$ for \ref{fig:lambdaG} and \ref{fig:lambdaD} respectively. Additionally, both plots contain the convex hull of the projections of the 3-D plot into the $\langle \hat{\lambda}\rangle$-$\langle \hat{G}\rangle$ and  $\langle \hat{W}\rangle$-$\langle \hat{G}\rangle$ planes. In Figure \ref{fig:lambdaG}, $\alpha$ and $\beta$ are marked to distinguish two regions of exciton condensation character.}
	\label{fig:color}
\end{figure*}

\begin{figure*}
	\centering
	\begin{subfigure}{.49\linewidth}
		\begin{overpic}[width = \linewidth,trim={7.25cm 9.2cm 8.05cm 11.1cm}, clip]{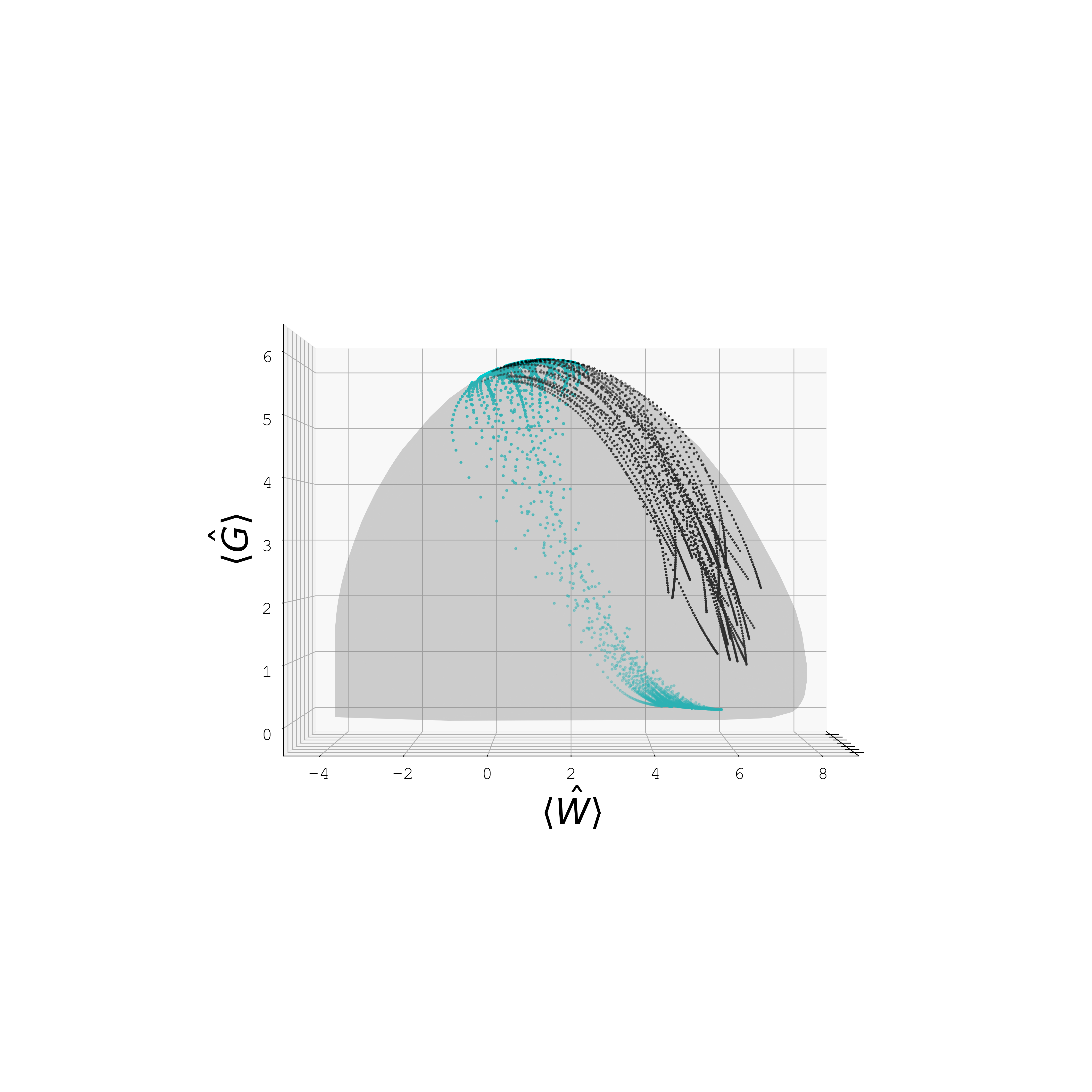}
			\put(-5,180){\normalsize(a)}
		\end{overpic}
		\phantomcaption
		\label{fig:4ptraj}
	\end{subfigure}
	\hfill
	\begin{subfigure}{.49\linewidth}
		\begin{overpic}[width = \linewidth,trim={7.25cm 9.2cm 8.05cm 11.1cm}, clip]{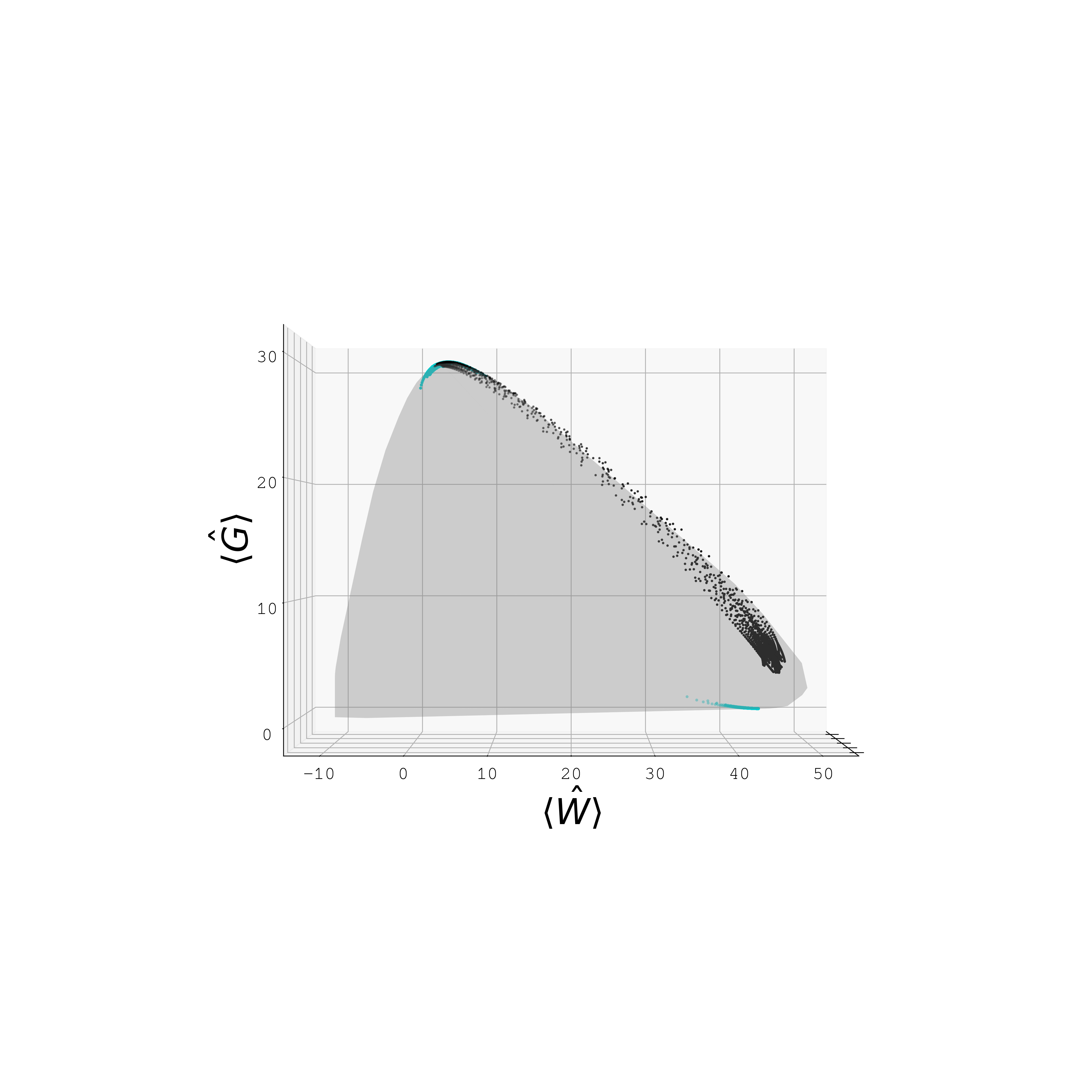}
			\put(-5,180){\normalsize(b)}
		\end{overpic}
		\phantomcaption
		\label{fig:10ptraj}
	\end{subfigure}
	\begin{subfigure}{.48\linewidth}
		\begin{overpic}[width = \linewidth,trim={.7cm 1cm 2.4cm 3.0cm}, clip]{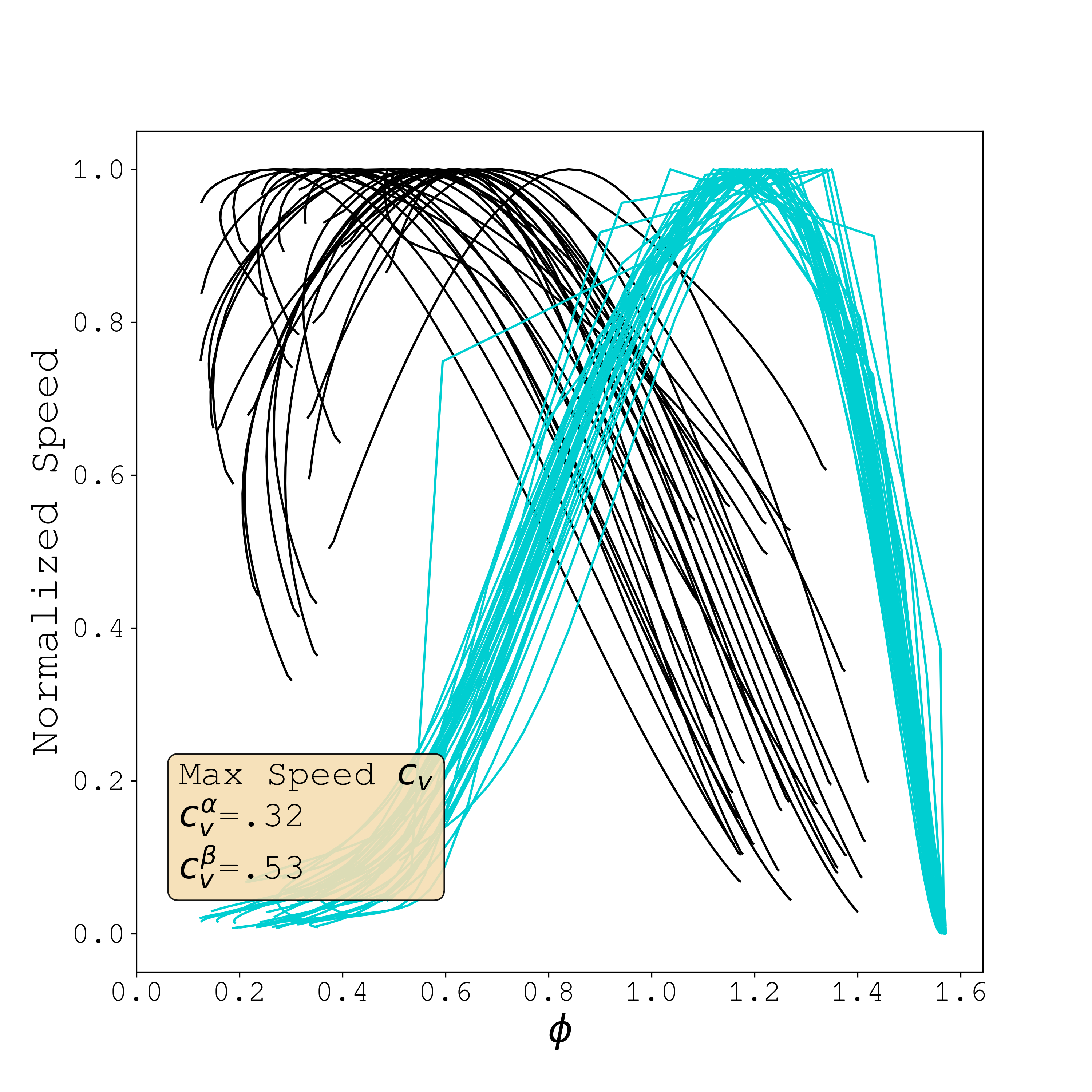}
			\put(-5,210){\normalsize(c)}
		\end{overpic}
		\phantomcaption
		\label{fig:4ptrajspeed}
	\end{subfigure}
	\hfill
	\begin{subfigure}{.48\linewidth}
		\begin{overpic}[width = \linewidth,trim={.7cm 1cm 2.4cm 3.0cm}, clip]{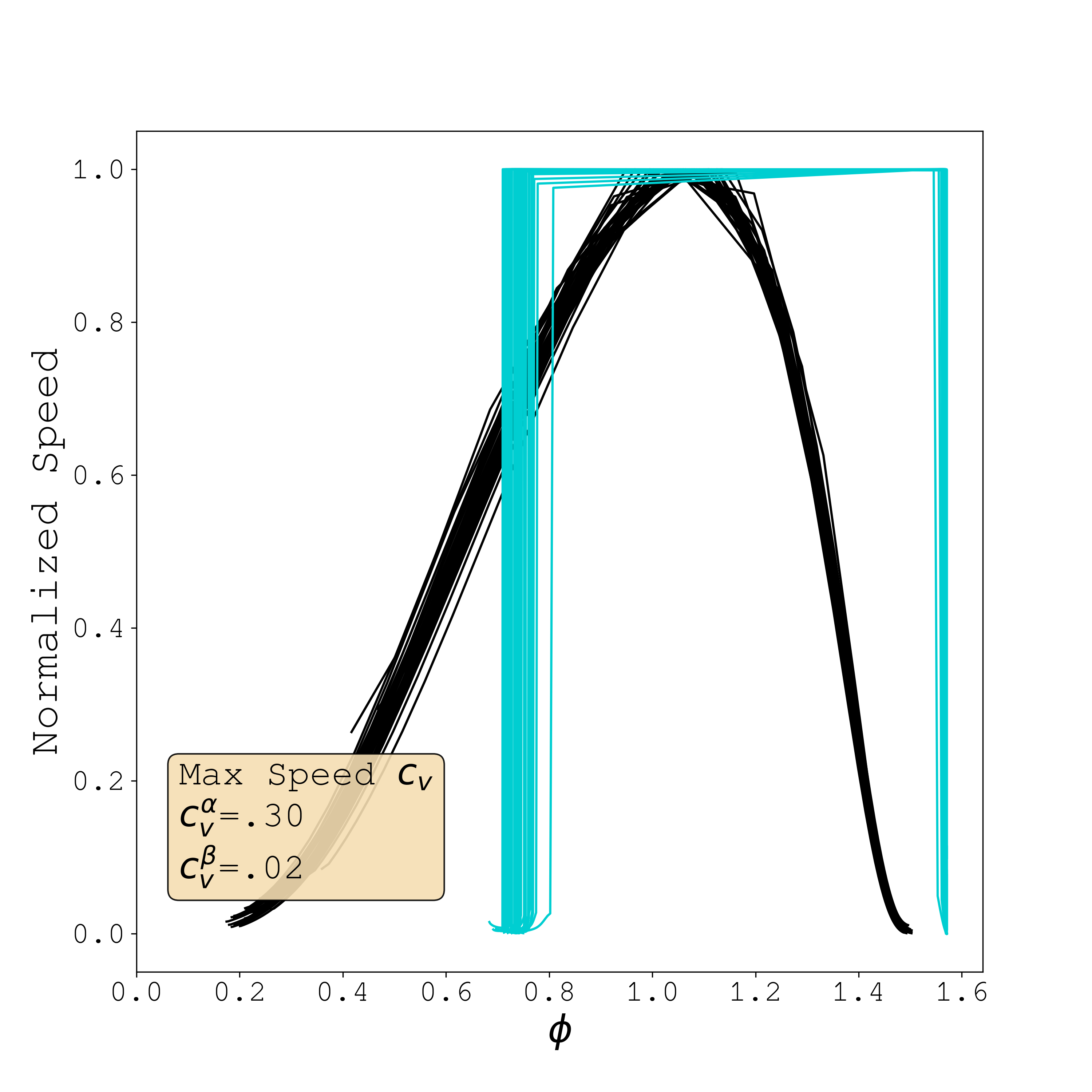}
			\put(-5,210){\normalsize(d)}
		\end{overpic}
		\phantomcaption
		\label{fig:10ptrajspeed}
	\end{subfigure}
	\vspace{-.5cm}
	\caption{Coalescing Trajectories from EC to FPC Regions. Figures \ref{fig:4ptraj} and \ref{fig:10ptraj} \tc{illustrate several dozen trajectories between randomly chosen Hamiltonian configurations in the parameter space outlined in Eq. \ref{eq:ham} from the $\alpha$, in black, and $\beta$, in turquoise, regions to the FPC region for N=4 and 10 particles, respectively}. A light grey shape illustrates the extent of the convex hull of the set of 2-RDMs. Figures \ref{fig:4ptrajspeed} and \ref{fig:10ptrajspeed} show the normalized 'speed' of the points along the convex set for both groups of trajectories with respect to the altitude angle $\phi$ in radians \tc{for N=4 and 10 particles, respectively}. The coefficient of variation \tc{,the ratio of the standard deviation to the mean or $\sigma/\mu$,} for the maximum speed attained along the trajectories is listed in the bottom left corner.}
	\label{fig:trajects}
\end{figure*}

\begin{figure*}
	\centering
%	\begin{subfigure}{.49\linewidth}
%		\begin{overpic}[width = \linewidth,trim={6cm 7.5cm 3.5cm 9.4cm}, clip]{4palphabeta.png}
%			\put(0,180){\normalsize(a)}
%		\end{overpic}
%		\phantomcaption
%		\label{fig:4palphabeta}
%	\end{subfigure}
%	\hfill
%	\begin{subfigure}{.49\linewidth}
%		\begin{overpic}[width = \linewidth,trim={6cm 7.5cm 3.5cm 9.4cm}, clip]{8palphabeta.png}
%			\put(0,180){\normalsize(b)}
%		\end{overpic}
%		\phantomcaption
%		\label{fig:8palphabeta}
%	\end{subfigure}
%	\hfill
	\begin{subfigure}{.49\linewidth}
%		\begin{overpic}[width = \linewidth,trim={1.3cm 1.8cm 1.5cm 4.5cm}, clip]{speed1.png}
		\begin{overpic}[width = \linewidth,trim={1.3cm 0cm 1.5cm 3cm}, clip]{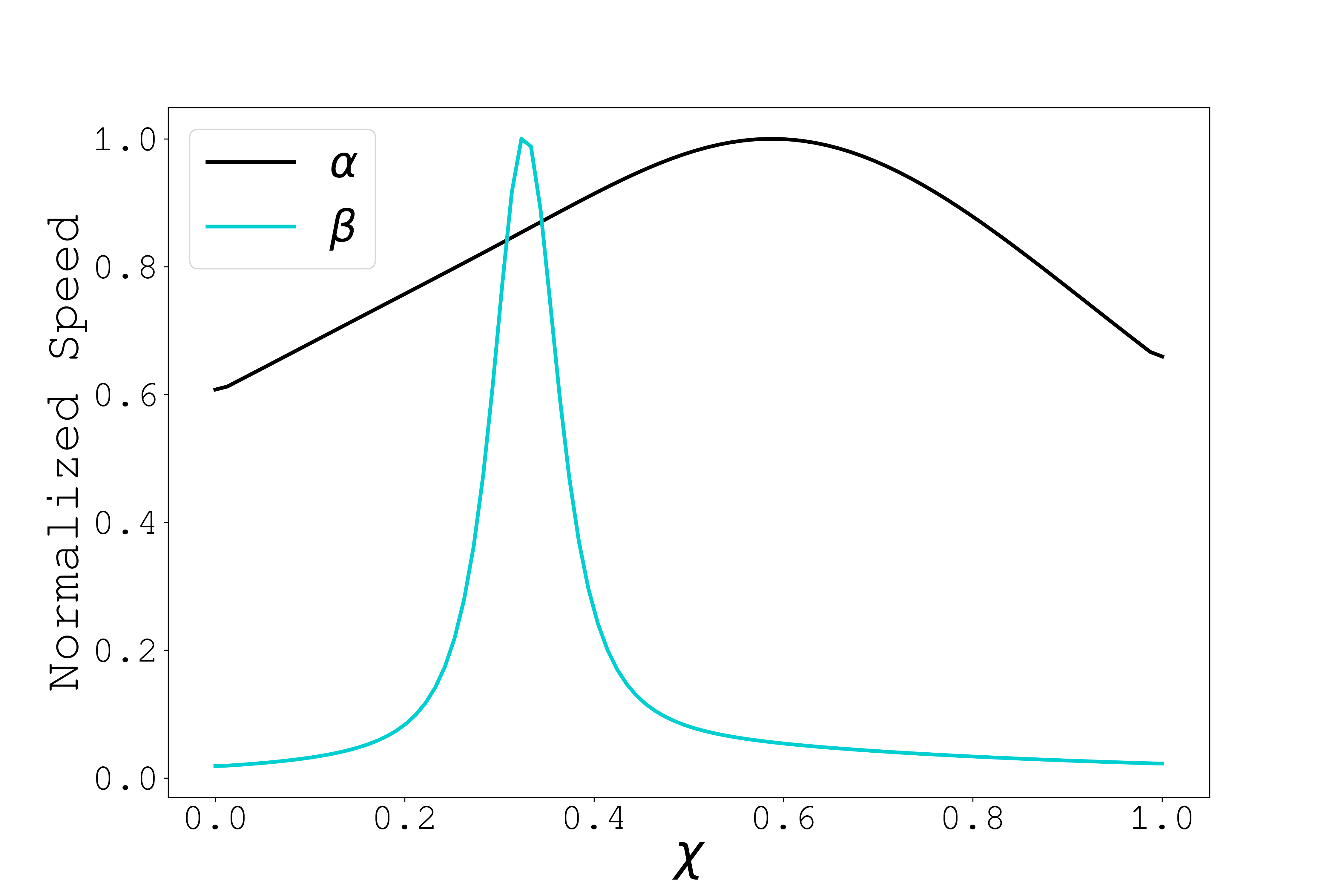}

			\put(-7,150){\normalsize(a)}
		\end{overpic}
		\phantomcaption
		\label{fig:speed4p}
	\end{subfigure}
	\begin{subfigure}{.49\linewidth}
		\begin{overpic}[width = \linewidth,trim={1.3cm 0cm 1.5cm 3cm}, clip]{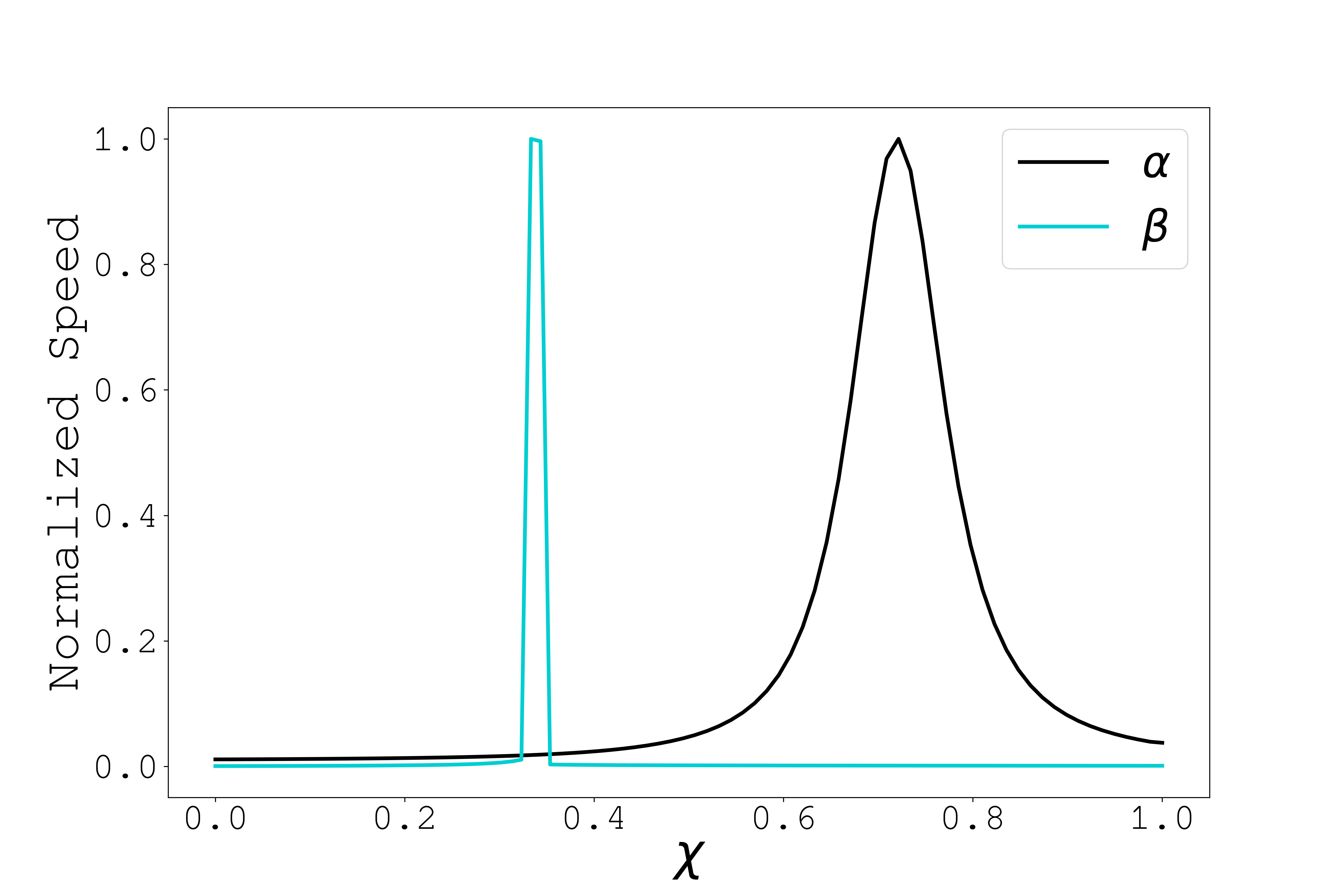}
%		\begin{overpic}[width = \linewidth,trim={1.3cm 1.8cm 1.5cm 4.5cm}, clip]{speed2.png}
			\put(-7,150){\normalsize(b)}
		\end{overpic}
		\phantomcaption
		\label{fig:speed10p}
	\end{subfigure}
	\vspace{-.5cm}
		\caption{Speed of the $\alpha$ and $\beta$ Trajectories. Figures \ref{fig:speed4p} and \ref{fig:speed10p} show the speed of a random trajectory from those plotted in Figure \ref{fig:trajects} for the 4 and 10 particle systems, respectively. $\chi$ is the parameter for the function describing the line between the starting and final Hamiltonians of the trajectories.}

	\label{fig:speed}
\end{figure*}

\begin{figure}
	\centering
	\begin{subfigure}{\linewidth}
		\begin{overpic}[width = \linewidth,trim={1.3cm 1.8cm 3.5cm 4.5cm}, clip]{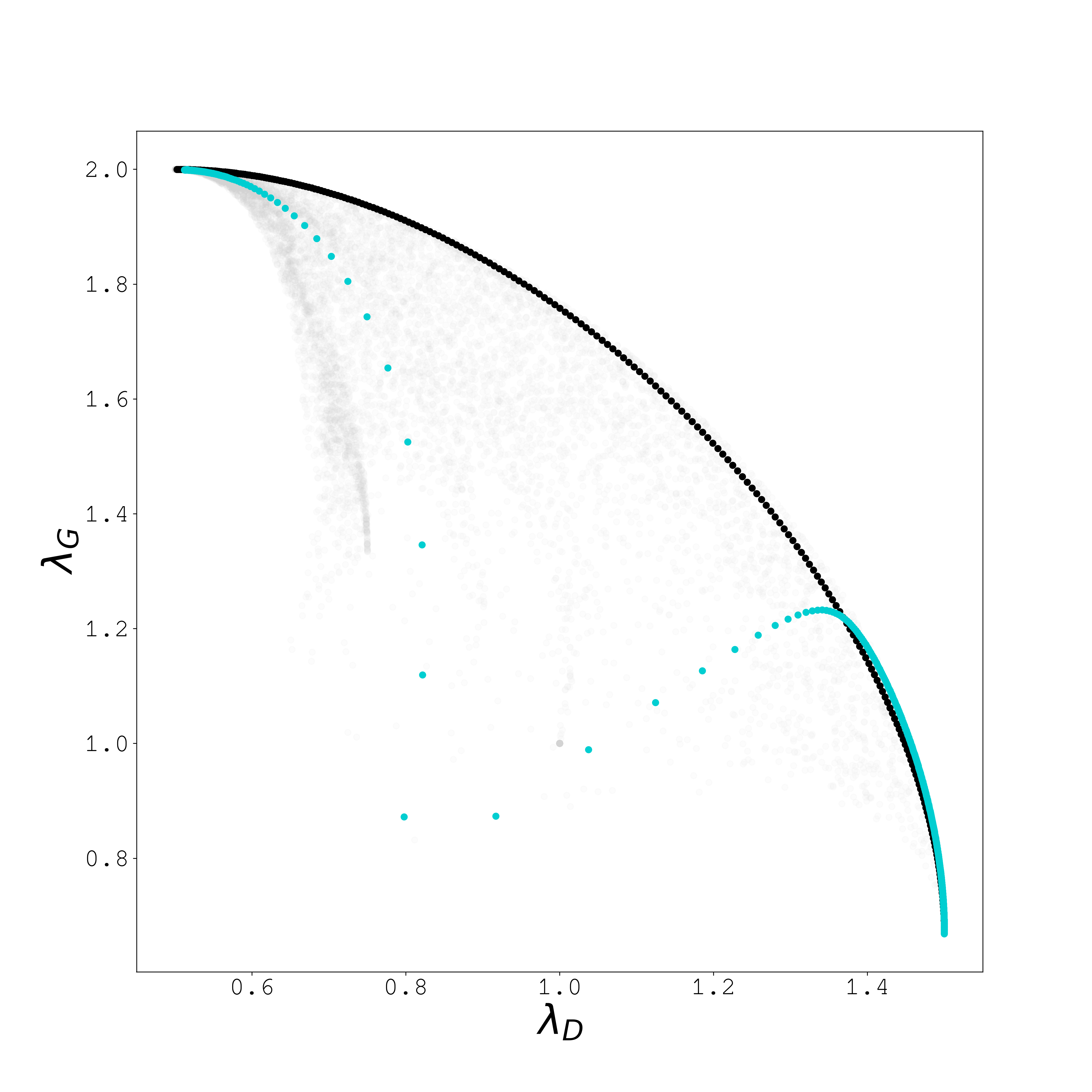}
			\put(-3,230){\normalsize(a)}
		\end{overpic}
		\phantomcaption
		\label{fig:4plam}
	\end{subfigure}
	\hfill
	\begin{subfigure}{\linewidth}
		\begin{overpic}[width = \linewidth,trim={1.3cm 1.8cm 3.5cm 4.5cm}, clip]{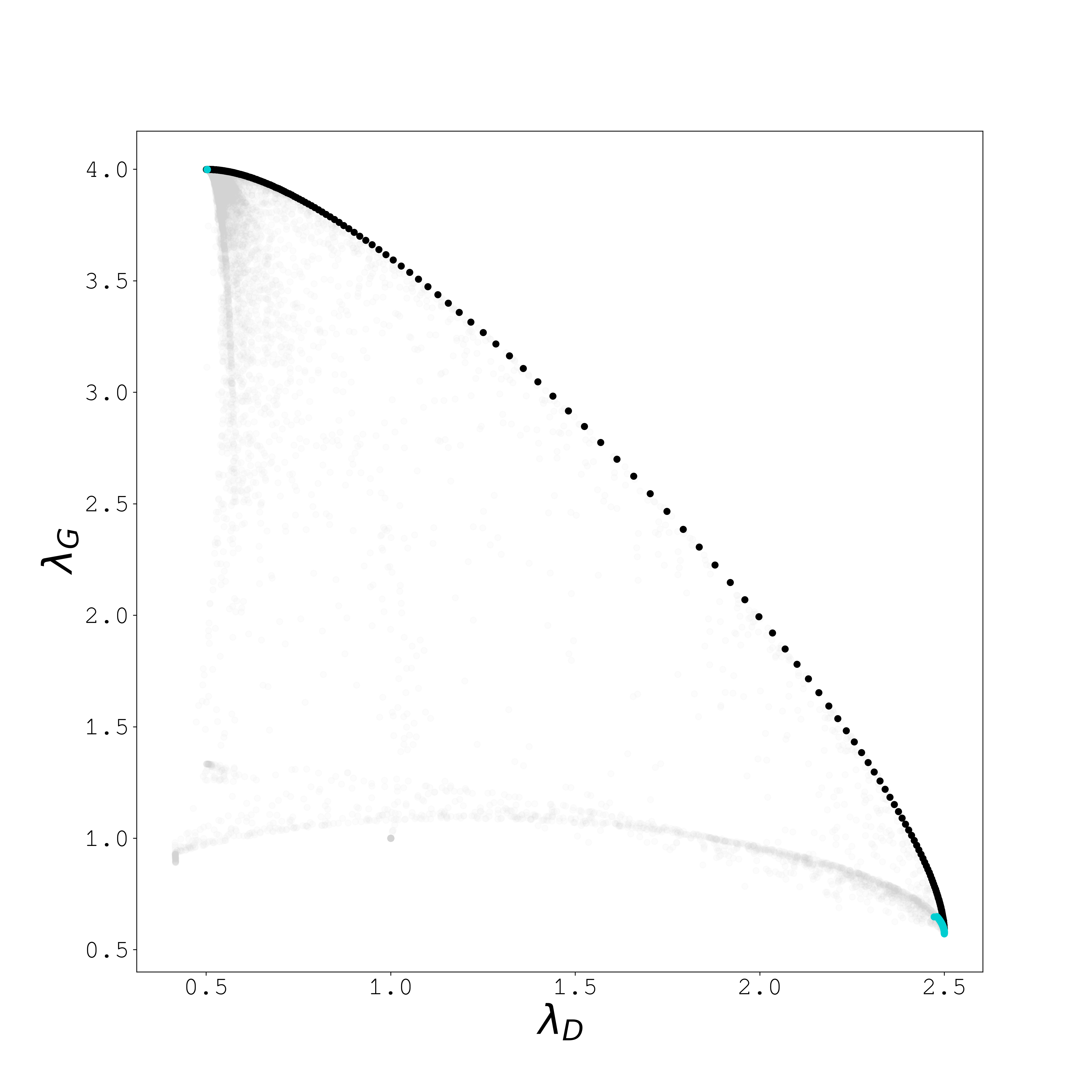}
			\put(-3,230){\normalsize(b)}
		\end{overpic}
		\phantomcaption
		\label{fig:8plam}
	\end{subfigure}
	\vspace{-.5cm}
	\caption{$\lambda$ Values along Trajectory. Figures \ref{fig:4plam} and \ref{fig:8plam} contain a $\alpha$, in black, and a $\beta$, in turquoise, trajectory for the 4 and 8 particle systems, respectively (the same as those in Fig \ref{fig:speed}). The light grey dots are the result of a random sampling of ground-state 2-RDMs. }
	\label{fig:lambdas}
\end{figure}

\begin{figure}
	\includegraphics[width=\linewidth]{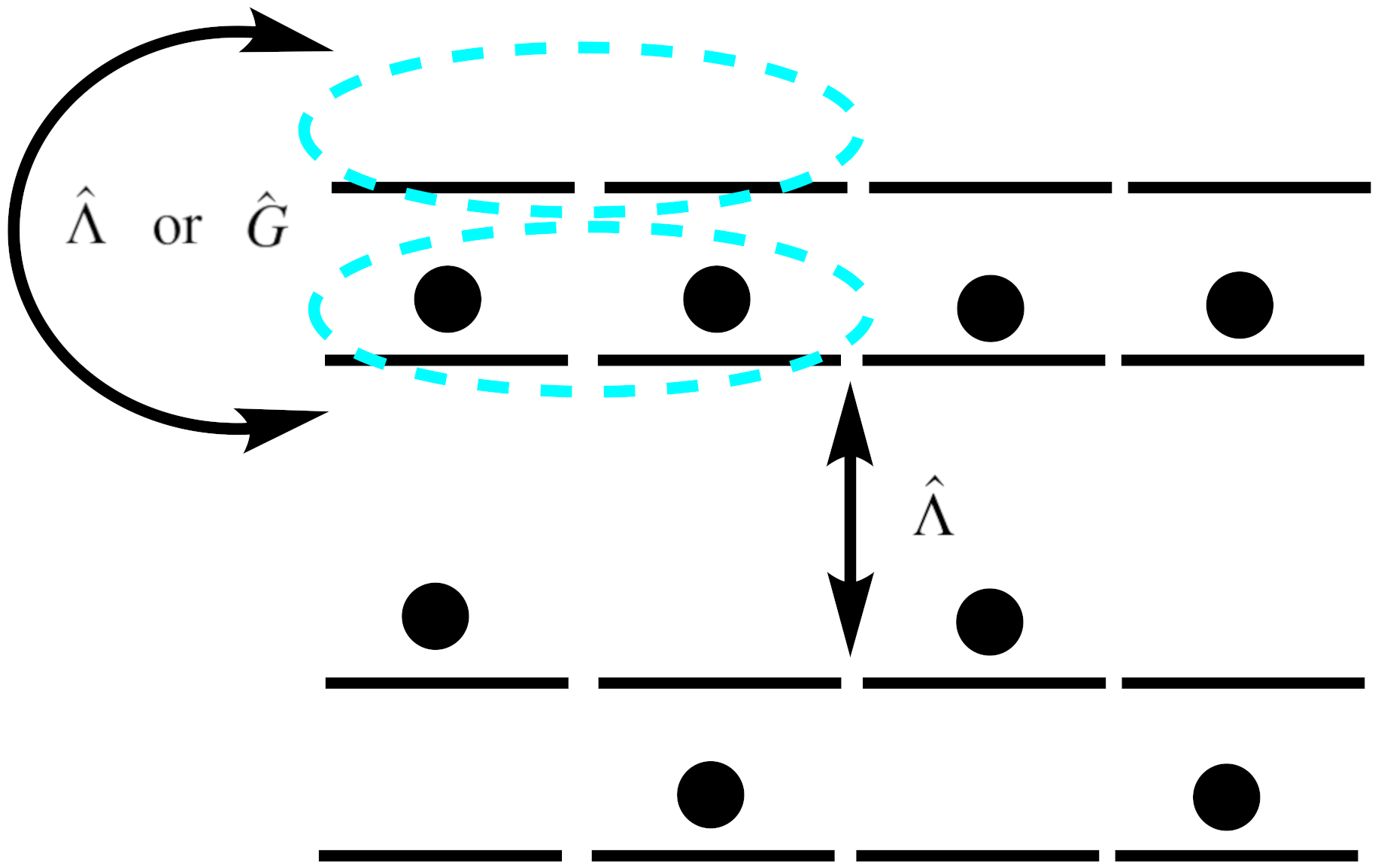}
	\caption{$\hat{\Lambda}$ Scattering. The upper most arrow shows a form of scattering that could be due to either the pair-force $\hat{G}$ term or the $\hat{\Lambda}$ term. While the lower arrow demonstrates how the $\hat{\Lambda}$ scattering is able to entangle states with large contributions to the maximal eigenstates of $\hat{W}$ with states that have large contributions to the maximal eigenstates of the pair-force term.}
	\label{fig:trans}
\end{figure}

\begin{figure}
	\centering
	\begin{subfigure}{\linewidth}
%			\begin{overpic}[width = \linewidth,trim={0cm 1.8cm 3.5cm 2cm}, clip]{alphaenergy.png}
			\begin{overpic}[width = \linewidth,trim={0cm 2.3cm 0cm 0cm}, clip]{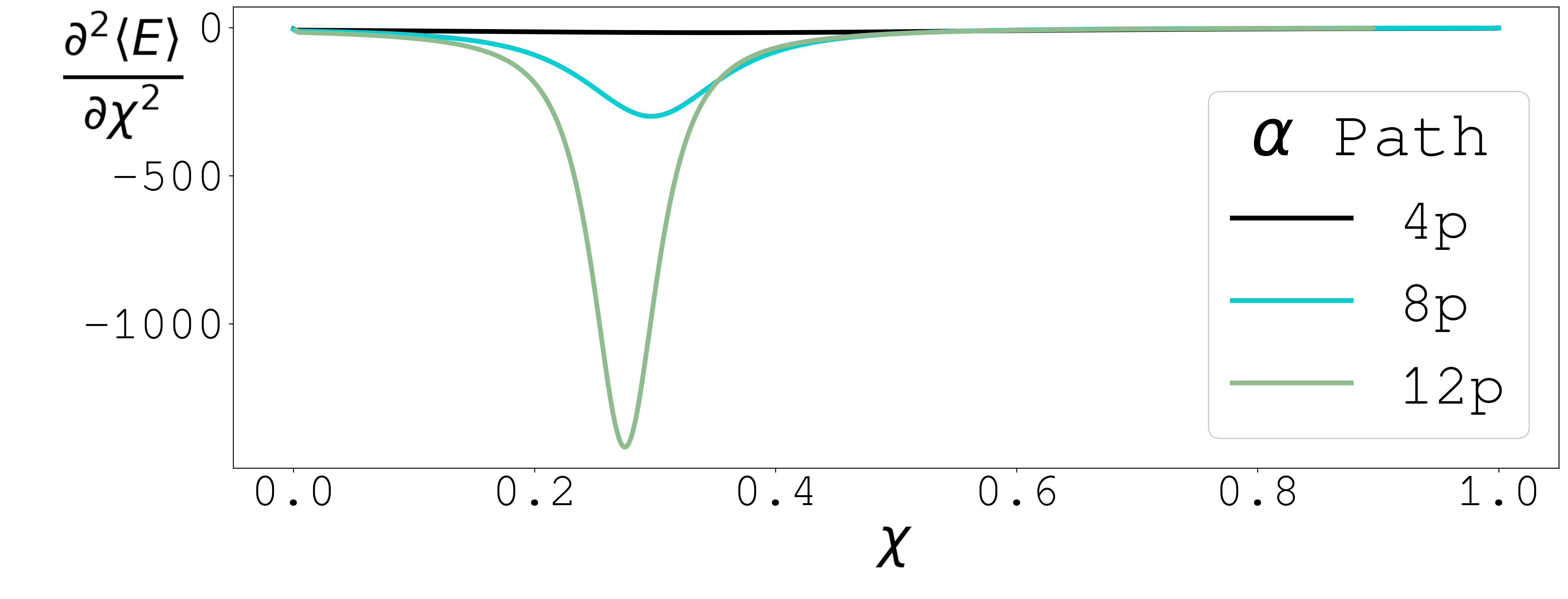}

			\put(0,100){\normalsize(a)}
		\end{overpic}
		\phantomcaption
		\label{fig:alphaenergy}
	\end{subfigure}
	\begin{subfigure}{\linewidth}
%		\begin{overpic}[width = \linewidth,trim={1.3cm 0cm 3.5cm 2cm}, clip]{betaenergy.png}
		\begin{overpic}[width = \linewidth,trim={0cm 0cm 0cm 0cm}, clip]{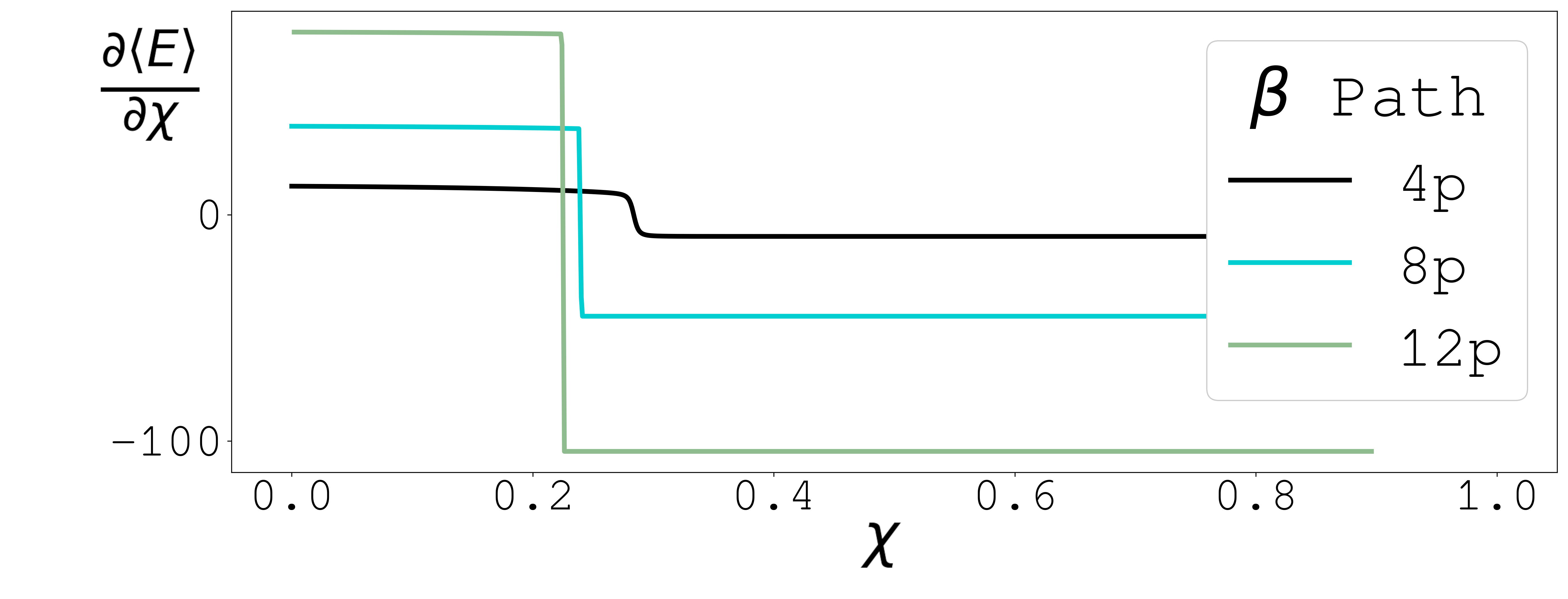}

			\put(0,100){\normalsize(b)}
		\end{overpic}
		\phantomcaption
		\label{fig:betaenergy}
	\end{subfigure}
	\vspace{-.5cm}
	\caption{Derivatives of the Ground State Energy Along Trajectory. Figure \ref{fig:alphaenergy} shows the second derivative of the ground-state energy with respect to $\chi$, for an $\alpha$ to FPC trajectory for the 4, 8, and 12 particle systems. Figure \ref{fig:betaenergy} shows the first derivative of the ground-state energy along a $\beta$ to FPC trajectory for the 4, 8, and 12 particle systems. These are the same trajectories as those shown in Fig. \ref{fig:speed}. }
	\label{fig:energy}
\end{figure}

\section{Results}

We address results attainable from finite-particle simulations of the system, and analyze these results through the RDM approach to QPTs in section~\ref{sec:R1}. These results then are used to extrapolate to the thermodynamic limit through construction of the ground-state wave functions in section~\ref{sec:R2}.

\subsection{Finite Particle}

\label{sec:R1}

		Investigation of the condensate character and critical behavior of the set of ground-state RDMs for the minimal four particle system can be seen in Figure~\ref{fig:color}. Figures~\ref{fig:lambdaD} and~\ref{fig:lambdaG} show the regions with exciton and fermion-pair condensation (FPC) or high values for $\lambda_D$ and $\lambda_G$, respectively. The regions with hotter coloring (red/orange) exhibit the highest levels of condensation, while cooler colors (blues) have limited to no condensation. Figure~\ref{fig:color} shows that there exists a significant yellow tinged area with high values of both $\lambda$ terms \tc{($\lambda_D,\lambda_G>1$)} between the regions of single condensation, indicating the existence of simultaneous fermion-exciton condensation (FEC) as reported in the initial and subsequent investigations of this system \cite{sager_simultaneous_2022,sager_potential_2020,sager_entangled_2022}. In Figure~\ref{fig:lambdaG}, two separate regions seem to maximize the exciton condensation (EC), marked as $\alpha$ and $\beta$, while one region in \ref{fig:lambdaD} maximizes the FPC. This RDM depiction of the system allows for a rapid assessment of the properties of the two different EC regions. Both EC regions have large positive values for the $\langle \hat{W}\rangle$, which is to be expected as the $\hat{W}$ term of the Hamiltonian scatters particle-hole pairs around the lattice. However, the two EC regions differ in regards to the expectation value of $\hat{\Lambda}$, which is the scattering term that moves pairs of particles (or holes) between the two degenerate energy levels. These observations raise several questions:  What is the critical behavior of the system as it transitions between the FPC and EC regions, and does this behavior differ when comparing the $\alpha$ or $\beta$ regions? Additionally, does the existence of two separate EC regions indicate the existence of two types of FECs?

As discussed in prior RDM studies, the `speed', Eq. \ref{eq:speed}, of a trajectory between two Hamiltonian configurations in the finite particle limit can serve as an indication of the presence of critical behavior in the thermodynamic limit \cite{gidofalvi_computation_2006,chen_geometry_2016}. For example, any abrupt changes in speed of the RDM along the surface of the convex set as the system is taken from the $\alpha$ to FPC region would indicate the development of some critical phenomenon between those regions. However, it is possible that the critical behavior along some of the paths between the regions differ drastically. Figure~\ref{fig:trajects} shows several dozen linear trajectories in the space of Hamiltonians, see Eq. \ref{eq:traj}. \tc{These trajectories are between randomly chosen Hamiltonian configurations in the Hamiltonian parameter space outlined in Eq. \ref{eq:ham}, where one end-point lies in the $\alpha$ region and the other in the fermion-pair condensate region}. Turquoise lines are used to indicate paths between the $\beta$ and FPC regions and black lines for trajectories between the $\alpha$ and FPC regions. For the four particle system, the trajectories in Figure~\ref{fig:4ptraj} show significant spread over the surface of the convex set of RDMs, which is \tc{depicted as a grey volume} in the figure. Additionally, the trajectories appear to differ in where they become sparse, or where their speed is the highest. This difference in the speed domain can be seen in Figure \ref{fig:4ptrajspeed}\tc{, which plots the speed of each of the trajectories with respect to the spherical coordinate altitude angle $\phi$.} The position of maximum speed differs significantly between the trajectories. The shapes of the speed curves also seem to differ, with a particularly large coefficient of variation, \tc{the standard deviation divided by the mean,} for the maximum speed of the $\beta$ curves.

However, as the particle number increases to 10 there is a decrease in the variability of the trajectories. The 3D-scatter plot of the curves shows a coalescing of the trajectories with the sparsity in the black $\alpha$ curves occurring in the same region, which is supported by the speed graph, Figure~\ref{fig:10ptrajspeed}, that displays that the maximum speed is obtained at nearly the same point for all of the curves. The $\beta$ curves also seem to become discontinuous at roughly the same point and their speed curves seem to be converging to a similar shape. The coefficient of variation of the maximum speed for the $\alpha$ curves, $c_v^\alpha$, is lower, and has been seen to decrease continually for higher particle numbers indicating that properties like the critical exponents are converging as well. The decrease in $c_v^\beta$, while more significant, has slightly less meaning as all of the curves become discontinuous.  This discontinuity reflects symmetry breaking in the system, and demonstrates the observation of abstract phenomena in the RDM formalism. Additionally, this data suggests that the critical behavior of the $\alpha$ and $\beta$ sets of trajectories can be roughly characterized by a single trajectory, and in the thermodynamic limit properties such as the critical exponents of the transition will converge to the same value. This convergence results from the dampening of the effect of local fluctuations on the expectation values of the ground-state RDMs away from critical points as the particle number increases \cite{sachdev_quantum_2011}.
%and demonstrates how the RDM formalism allows for visualization of abstract phenomenon.

The speeds for a single $\alpha$ trajectory are plotted in Figures~\ref{fig:speed4p} and \ref{fig:speed10p}. In these Figures, the hump showing the speed grows with the particle number, which indicates that it will likely become discontinuous in the thermodynamic limit. Therefore, this is the finite-particle signature of a quantum phase transition along this trajectory. The exact order of the QPT lying along the $\alpha$ trajectories cannot be determined from the speed alone, but can be discovered through more a traditional analysis of the ground-state energy. The turquoise or $\beta$ line shows this same acceleration to a much larger degree. The RDMs along the $\beta$ line seem to ``leap" from the region of exciton condensation to fermion condensation almost instantaneously in the eight-particle case. This leap, as discussed before, is an indication of symmetry breaking, and that this transition is first order.

This ``leap" can also be seen in Figure~\ref{fig:lambdas} which shows the same trajectories, but now in the space of the signatures of condensation for excitons and particle-particle pairs, i.e., a plot of $\lambda_G$ vs $\lambda_D$. This plot contains a light-grey scattering of a random sample of RDMs to illustrate the extent of the accessible $\lambda$ region. In the four- and eight-particle cases the black line travels through the space of RDMs along an ellipse that simultaneously maximizes both of the $\lambda$ values. However, the turquoise line for the four-particle case shows a drastic drop in the eigenvalues along the trajectory, while the eight-particle case skips the FEC region entirely. This demonstrates that the FEC only exists along the $\alpha$ trajectory, and that only one type of FEC exists in the thermodynamic limit of the system. These results for the $\beta$ trajectory suggest that the large positive energy contribution of the $\hat{\Lambda}$ term in the Hamiltonian directly conflicts with the system's ability to become a FPC. This is a result of two separate functions of this scattering term. The first can be observed in Figure~\ref{fig:lambdaG}, where the FPC condensate region seems to lie mostly above $\langle\hat{\Lambda}\rangle$=0 as seen in the projection into the $\langle \hat{\Lambda}\rangle-\langle\hat{G}\rangle$ plane. This preference for FPC to lie above $\langle\hat{\Lambda}\rangle$=0 is likely the result of the $\hat{\Lambda}$ scattering working cooperatively with the $\hat{G}$ pair-force term to lower the energy, by moving pairs of particles between the paired orbitals in the same way as the pair-force term as illustrated in the upper half of Figure~\ref{fig:trans}. The potentially more crucial function of the $\hat{\Lambda}$ term, however, is the connection of states with large contributions to the $\hat{W}$ term with the pair-force states. In other words, the $\hat{\Lambda}$ scattering provides off-diagonal entanglement between exciton states and fermion-pair states as seen in the bottom half of Figure~\ref{fig:trans}. When this scattering is energetically unfavorable, as is seen in the $\beta$ trajectories in Figure~\ref{fig:lambdas}, there is a dip or a leap in the eigenvalues as the system is unable to entangle the states from the two regions of condensation, resulting in what appears to be the formation of a first-order phase transition between the states which minimize $\hat{W}$ and $\hat{G}$.

Analysis of the ground-state energy of the system as it travels along the trajectories outlined above can be seen in Figure~\ref{fig:energy}. These results largely confirm those already gathered from RDM techniques, but allow for a deeper analysis of the order of the phase transitions postulated. Figure~\ref{fig:alphaenergy} shows that the second derivative of the energy along the $\alpha$ trajectory is rapidly growing as the particle number increases, likely becoming discontinuous in the thermodynamic limit. This indicates that a second-order or higher-order quantum phase transition is occurring between the $\alpha$ and FPC regions of condensation. While along the $\beta$ trajectories, the first derivative of the energy is seemingly already discontinuous even with a finite number of particles, which is a result of the actual level crossing that occurs in the system \tc{(lower order derivatives of the energy can be found in the Supplemental Material~\footnote{See Supplemental Material at
[URL will be inserted by publisher] for ground-state energies and their derivatives along the quantum phase transitions.})}. This means that a first-order phase transition exists between the $\beta$ and FPC regions, but this was already clear from the discontinuity of the trajectories on the RDM set.

\subsection{Thermodynamic Limit}

\label{sec:R2}

The first-order transition can be extrapolated to the thermodynamic limit through analysis of the structure of the ground-state wave functions. The $\beta$ region Hamiltonians are dominated by contributions of $\hat{\Lambda}-\hat{W}$. By diagonalizing this term in the finite-particle limit, a clear pattern emerges in the ground-state subspace. The ground-state is doubly degenerate, and this subspace can be characterized by wave functions of the form:
\begin{equation}
	\begin{split}
		\ket{\Psi_1^\beta} &= \left( -1 \right)^{\hat{L}} \Biggl[ c_1 \biggl(\sum_j^{N/2} \sum_i \ket{\phi_i^{2j}} \biggr) \\
		 &+ c_2 \biggl( \sum_j^{N/2-1} \sum_i \ket{\phi_i^{2j+1}} \biggr) \Biggr]
	\end{split}
\end{equation}
\begin{equation}
	\begin{split}
		\ket{\Psi_2^\beta} &= \left( -1 \right)^{\hat{L'}} \Biggl[ c_2 \biggl(\sum_j^{N/2} \sum_i \ket{\phi_i^{2j}} \biggr) \\
		 &+ c_1 \biggl( \sum_j^{N/2-1} \sum_i \ket{\phi_i^{2j+1}} \biggr) \Biggr]
	\end{split}
\end{equation}
Where $\ket{\phi_i^{k}}$ is the $i^{th}$ Lipkin-like wave function (a state without occupations of both $p$ and $p+N$ orbitals) with $k$ occupied orbitals in the `upper' energy level (or $k$ occupied orbitals numbered greater than $N$) and
\begin{equation}
	\hat{L} = \lfloor\left(\sum_i^N {\hat a}^\dagger_i{\hat a}_i\right)/2\rfloor
	\label{eq:L}
\end{equation}
\begin{equation}
	\hat{L'} = \lfloor\left(\sum_i^N {\hat a}^\dagger_i{\hat a}_i+1\right)/2\rfloor,
\end{equation}
Which are equivalent to counting the number of pairs of particles in the `upper' (p $\geq$ N) energy level. It is apparent from the structure of the wave function that it represents an exciton condensate because of the large variability in both the number and location of excitations throughout the lattice.

The proposed first-order transition occurs along the path $\hat{H}=\hat{\Lambda}-\hat{W}\rightarrow\hat{G}$. However, it can be shown (see Appendix A) that $\hat{G}\ket{v}=0$ where $v\in \ket{\Psi_1^\beta}\otimes\ket{\Psi_2^\beta}$. This means that $\ket{v}$ is an eigenvalue for any Hamiltonian along the path. Therefore the ground-state will only change at an actual energy level crossing, as the degenerate subspace is diagonal, and thus uncoupled, to any other energy levels. Slight perturbations to the Hamiltonian from the path outlined above, only break the degeneracy of the ground-state subspace, but do not change its diagonal and uncoupled nature.

Similar analysis of the ground-states in the $\alpha$ and FPC regions reveal wave functions of the form:
\begin{equation}
	\begin{split}
		\ket{\Psi_1^\alpha} &= \Biggl[ c_1 \biggl(\sum_j^{N/2} \sum_i \ket{\phi_i^{2j}} \biggr) \\
		 &+ c_2 \biggl( \sum_j^{N/2-1} \sum_i \ket{\phi_i^{2j-1}} \biggr) \Biggr]
	\end{split}
\end{equation}
\begin{equation}
	\begin{split}
		\ket{\Psi_2^\alpha} &= \Biggl[ c_2 \biggl(\sum_j^{N/2} \sum_i \ket{\phi_i^{2j}} \biggr) \\
		 &- c_1 \biggl( \sum_j^{N/2-1} \sum_i \ket{\phi_i^{2j-1}} \biggr) \Biggr]
	\end{split}
\end{equation}
and
\begin{equation}
	\ket{\Psi_{FPC}}= N\left(\sum_i \ket{\psi_i}\right)
\end{equation}
where $\ket{\psi_i}$ is the $i^{\rm th}$ BCS-like state, meaning that if orbital $2j$ is occupied, then orbital $2j-1$ must be occupied as well. The $\alpha$ wave functions are not eigenvectors of $\hat{G}$ due to their internal parity relying on even or odd occupation of the different energy levels. However, $\hat{G}$ acting on $\alpha$ wave functions generates non-Lipkin-like states (see Appendix A for the action of $\hat{G}$ on Lipkin-like states), a subset of which are BCS-like states. This coupling between the two different ground-states, is what allows for a smooth transition (or avoided level crossing) between the regions of the RDM. This further confirms that if a QPT exists in the thermodynamic limit (which as discussed in earlier sections looks likely as the finite particle signature of the transition seems to grow sharper with increasing particle number), it will be a 2nd order or higher transition.

\section{Discussion and Conclusions}

The exploration of the critical behavior of this system brings to light several interesting revelations. Through a combination of RDM techniques, a mapping of the space of Hamiltonians to the convex set of RDMs reveals multiple excitonic phases in the system. This color map also highlights the regions with FEC states. Increasing the particle number increases the curvature of the set, or increases the `speed' of the trajectories traveling through the regions containing the FEC states, resulting in the discovery of a first-order quantum phase transition between the ``$\beta$" exciton and the fermion-fermion condensate regions, and a second-order transition between the ``$\alpha$'' exciton and FPC regions. Using the RDM set as a guide, it is possible to apply more traditional wave function analysis to the transitions. The characteristic ground-state wave functions for each of the regions are then used to extrapolate the behavior of the system at the thermodynamic limit confirming the existence of the QPTs.

Several interesting questions still remain about the system. The first concerns the behavior of the system at non-zero temperature. How resilient are the condensates to temperature, and do novel phases emerge? Additionally, would constraining the BCS and LMG terms to a limited spatial area, to make the interactions more applicable to common crystal systems, completely suppress condensation? This would require defining a specific lattice shape, and with the BCS term it might be possible to generate zero energy edge states in the system much like the Kitaev chain. As two different symmetries, $L$ and even/odd parity, seem to dominate the system, perhaps these could be used to generate symmetry protected topological states.

An experimental realization of this material may be able to take advantage of the finding that the FEC region lies between fermion-fermion pair and exciton condensate phases. This suggests that one route to an FEC material is placing a excitonic material, like a bilayered system or a gapped system \cite{bouscher_enhanced_2022}, on a bulk superconductor. Ideally, this bilayered system would also be able to exhibit superconductivity in some regimes (a possible candidate being twisted graphene). The superconducting character of the heterogeneous system could then be potentially controlled by varying the temperature of the system or by inducing a voltage along the junction of the materials. By lowering the temperature or increasing the voltage, the bulk superconductor will begin to donate a greater number of Cooper pairs into the bilayer, and potentially increase the superconducting character of the bilayer until it exhibits FEC. Studying simple composite systems computationally will increase our theoretical understanding of this novel form of condensation and help build the framework necessary for an experimental realization of Fermion-Exciton Condensation.

%\noindent {\bf Supplemental Material:} Ground-state energies and their derivatives are reported along the quantum phase transitions.

\section{Acknowledgements}
D.A.M. gratefully acknowledges the Department of Energy, Office of Basic Energy Sciences Grant No. DE-SC0019215 and the National Science Foundation Grant No. CHE-2155082.

\appendix

\tc{\section{Thermodynamic Limit}}

The proposed $1^{st}$ order QPT arises from the symmetry of the degenerate ground-state in the $\beta$ exciton region. This symmetry, $L$ (Eq. \ref{eq:L}), which swaps the sign of states with `upper' (orbital number $\geq N$) occupation values differing by 2, preserves the degenerate subspace in the FPC region, dominated by $\hat{G}$. This can be proven by differentiating the basis functions given in the paper, with new quantum numbers, which allow for an easier analysis of the effect of $\hat{G}$ on the wave functions.

With these new quantum numbers, $\ket{\Psi^\beta_1}$ can be defined as:
\begin{equation}
	\begin{split}
		\ket{\Psi} &= c_1 \left( \sum_{j=0,2,..}^{N/2} \sum_{i=0,2,..}^{N/2-j} \ket{i,j}_L - \ket{i+1,j}_L \right) \\
		&+ c_2 \left( \sum_{j=1,3,..}^{N/2} \sum_{i=0,2,..}^{N/2-j} \ket{i,j}_L - \ket{i+1,j}_L\right)
	\end{split}
\end{equation}
where $\ket{i,j}_L$ is a Lipkin-like wave function (i.e. orbitals $p$ and $p+N$ cannot be occupied simultaneously), $i$ is the number of occupied adjacent orbitals $\left(2p,2p-1\right)$ where $p\geq N$, and $j$ is the number of Lipkin-like pairs where $\left(2p,2p-1+N\right)$ or $\left(2p+N,2p-1\right)$ are occupied. It should be noted that the first and second term contain wave functions with even and odd occupations of the `upper' energy level, respectively. For any $i$ and $j$, there are $\binom{N/2}{j}\binom{N/2-j}{i}$ arrangements of particles in the lattice which satisfy these conditions, and $\ket{i,j}_L$ is an equal and positive superposition of all of these configurations.

Recall that $\hat{G}$ is only able to move pairs of particles occupying $(2k,2k-1)$ orbitals, therefore it cannot act on particles forming Lipkin-like pairs. This means $j$ is unaffected by the BCS operator. Applying $\hat{G}$ to a $\ket{i,j}_L$ results in a new superposition:
\begin{equation}
	\begin{split}
		&\hat{G}\ket{i,j}_L = G_L^=\ket{i,j}_L+G_L^+\ket{i+1,j}_L+G_L^-\ket{i-1,j}_L \\
		&+ G_{NL}^=\ket{i,j}_{NL}+G_{NL}^+\ket{i+1,j}_{NL}+G_{NL}^-\ket{i-1,j}_{NL}
	\end{split}
\end{equation}
Where $\ket{i,j}_{NL}$ is a state with one set of occupied orbitals $\left(2p,2p-1,2p+N,2p-1+N\right)$, which is a state not found in the traditional LMG model. These Non-Lipkin-like conditions result in $\binom{N/2}{j}\binom{N/2-j}{1}\binom{N/2-j-1}{i-1}\binom{N/2-j-i}{1}$ possible arrangements of particles in the lattice which are also summed in an equal superposition to generate the wave function $\ket{i,j}_{NL}$. It can be rationalized that the $G$ terms are determined by the following relationship:
\begin{equation}
	G_{L/NL}^{+/-/=}=\frac{D\left(\ket{i,j}_L\right)\left(E_{L/NL}^{+/-/=} \right)}{D\left(\ket{i\pm1/i,j}_{L/NL}\right)}
\end{equation}
where $D\left(\ket{i,j}\right)$ is the number of arrangements of particles in the lattice which satisfy $i$ and $j$ (given above), and $E$ is the number of arrangements any specific configuration that satisfies $i$ and $j$ can be excited/de-excited to. $E$ must be the same for every arrangement of the same $i$ and $j$. It is easy to rationalize this fact for Lipkin states going to Lipkin states as the number of excitations/de-excitations is exactly equal to the number of BCS pairs in the `lower'/`upper' energy levels (changing the occupied orbital label by $\pm N$).  This number, $i$ or $N-i-j$, must be the same for any of the arrangements of particles in the lattice, which compose $\ket{i,j}_L$.

Solving this equation for the various values of $G$ reveals:
\begin{equation}
	\begin{split}
		\hat{G}\ket{i,j}_L&=\left(N-j\right)\ket{i,j}_L \\
		&+\left(i+1\right)\ket{i+1,j}_L \\
		&+\left(N-j-i+1\right)\ket{i-1,j}_L \\
		&+2\ket{i,j}_{NL} \\
		&+\ket{i+1,j}_{NL} \\
		&+\ket{i-1,j}_{NL} .
	\end{split}
\end{equation}
Finally applying these terms to $G$ acting on $\ket{\Psi}$:
\begin{equation}
	\begin{split}
		&\hat{G}\ket{\Psi}=\hat{G}\left(..-\ket{i+1,j}_L+\ket{i,j}_L-\ket{i-1,j}_L\right)\\
		&=..+\bigl(-\left(N-j-(i+1)+1\right)+\left(N-j\right)-i\bigr)\ket{i,j}_L\\
		&+\left(-1+2-1\right)\ket{i,j}_{NL}+...=0
	\end{split}
\end{equation}
demonstrates that $\ket{\Psi_1^\beta}$ is an eigenvalue of $\hat{G}$. The exact same arguments can be applied to $\ket{\Psi_2^\beta}$, and therefore $\hat{G}\ket{v}=0$.

\bibliography{Phase}

\begin{thebibliography}{59}%
\makeatletter
\providecommand \@ifxundefined [1]{%
 \@ifx{#1\undefined}
}%
\providecommand \@ifnum [1]{%
 \ifnum #1\expandafter \@firstoftwo
 \else \expandafter \@secondoftwo
 \fi
}%
\providecommand \@ifx [1]{%
 \ifx #1\expandafter \@firstoftwo
 \else \expandafter \@secondoftwo
 \fi
}%
\providecommand \natexlab [1]{#1}%
\providecommand \enquote  [1]{``#1''}%
\providecommand \bibnamefont  [1]{#1}%
\providecommand \bibfnamefont [1]{#1}%
\providecommand \citenamefont [1]{#1}%
\providecommand \href@noop [0]{\@secondoftwo}%
\providecommand \href [0]{\begingroup \@sanitize@url \@href}%
\providecommand \@href[1]{\@@startlink{#1}\@@href}%
\providecommand \@@href[1]{\endgroup#1\@@endlink}%
\providecommand \@sanitize@url [0]{\catcode `\\12\catcode `\$12\catcode
  `\&12\catcode `\#12\catcode `\^12\catcode `\_12\catcode `\%12\relax}%
\providecommand \@@startlink[1]{}%
\providecommand \@@endlink[0]{}%
\providecommand \url  [0]{\begingroup\@sanitize@url \@url }%
\providecommand \@url [1]{\endgroup\@href {#1}{\urlprefix }}%
\providecommand \urlprefix  [0]{URL }%
\providecommand \Eprint [0]{\href }%
\providecommand \doibase [0]{https://doi.org/}%
\providecommand \selectlanguage [0]{\@gobble}%
\providecommand \bibinfo  [0]{\@secondoftwo}%
\providecommand \bibfield  [0]{\@secondoftwo}%
\providecommand \translation [1]{[#1]}%
\providecommand \BibitemOpen [0]{}%
\providecommand \bibitemStop [0]{}%
\providecommand \bibitemNoStop [0]{.\EOS\space}%
\providecommand \EOS [0]{\spacefactor3000\relax}%
\providecommand \BibitemShut  [1]{\csname bibitem#1\endcsname}%
\let\auto@bib@innerbib\@empty
%</preamble>
\bibitem [{\citenamefont {Onnes}(1913)}]{onnes_investigations_1913}%
  \BibitemOpen
  \bibfield  {author} {\bibinfo {author} {\bibfnamefont {H.~K.}\ \bibnamefont
  {Onnes}},\ }\bibfield  {title} {\bibinfo {title} {Investigations into the
  properties of substances at low temperatures, which have led, amongst other
  things, to the preparation of liquid helium}\ }(\bibinfo {year}
  {1913})\BibitemShut {NoStop}%
\bibitem [{\citenamefont {Kapitza}(1938)}]{kapitza_viscosity_1938}%
  \BibitemOpen
  \bibfield  {author} {\bibinfo {author} {\bibfnamefont {P.}~\bibnamefont
  {Kapitza}},\ }\bibfield  {title} {\bibinfo {title} {Viscosity of {Liquid}
  {Helium} below the $\lambda$-{Point}},\ }\href
  {https://doi.org/10.1038/141074a0} {\bibfield  {journal} {\bibinfo  {journal}
  {Nature}\ }\textbf {\bibinfo {volume} {141}},\ \bibinfo {pages} {74}
  (\bibinfo {year} {1938})}\BibitemShut {NoStop}%
\bibitem [{\citenamefont {Hartnoll}\ \emph {et~al.}(2008)\citenamefont
  {Hartnoll}, \citenamefont {Herzog},\ and\ \citenamefont
  {Horowitz}}]{hartnoll_building_2008}%
  \BibitemOpen
  \bibfield  {author} {\bibinfo {author} {\bibfnamefont {S.~A.}\ \bibnamefont
  {Hartnoll}}, \bibinfo {author} {\bibfnamefont {C.~P.}\ \bibnamefont
  {Herzog}},\ and\ \bibinfo {author} {\bibfnamefont {G.~T.}\ \bibnamefont
  {Horowitz}},\ }\bibfield  {title} {\bibinfo {title} {Building a {Holographic}
  {Superconductor}},\ }\href {https://doi.org/10.1103/PhysRevLett.101.031601}
  {\bibfield  {journal} {\bibinfo  {journal} {Physical Review Letters}\
  }\textbf {\bibinfo {volume} {101}},\ \bibinfo {pages} {031601} (\bibinfo
  {year} {2008})}\BibitemShut {NoStop}%
\bibitem [{\citenamefont {Anderson}(2013)}]{anderson_twenty-five_2013}%
  \BibitemOpen
  \bibfield  {author} {\bibinfo {author} {\bibfnamefont {P.~W.}\ \bibnamefont
  {Anderson}},\ }\bibfield  {title} {\bibinfo {title} {Twenty-five {Years} of
  {High}-{Temperature} {Superconductivity} – {A} {Personal} {Review}},\
  }\href {https://doi.org/10.1088/1742-6596/449/1/012001} {\bibfield  {journal}
  {\bibinfo  {journal} {Journal of Physics: Conference Series}\ }\textbf
  {\bibinfo {volume} {449}},\ \bibinfo {pages} {012001} (\bibinfo {year}
  {2013})}\BibitemShut {NoStop}%
\bibitem [{\citenamefont {Chen}\ \emph {et~al.}(2020)\citenamefont {Chen},
  \citenamefont {Jiang}, \citenamefont {Zhang}, \citenamefont {Liu},
  \citenamefont {Liu}, \citenamefont {Wang},\ and\ \citenamefont
  {Wang}}]{chen_atomic_2020}%
  \BibitemOpen
  \bibfield  {author} {\bibinfo {author} {\bibfnamefont {C.}~\bibnamefont
  {Chen}}, \bibinfo {author} {\bibfnamefont {K.}~\bibnamefont {Jiang}},
  \bibinfo {author} {\bibfnamefont {Y.}~\bibnamefont {Zhang}}, \bibinfo
  {author} {\bibfnamefont {C.}~\bibnamefont {Liu}}, \bibinfo {author}
  {\bibfnamefont {Y.}~\bibnamefont {Liu}}, \bibinfo {author} {\bibfnamefont
  {Z.}~\bibnamefont {Wang}},\ and\ \bibinfo {author} {\bibfnamefont
  {J.}~\bibnamefont {Wang}},\ }\bibfield  {title} {\bibinfo {title} {Atomic
  line defects and zero-energy end states in monolayer {Fe}({Te},{Se})
  high-temperature superconductors},\ }\href
  {https://doi.org/10.1038/s41567-020-0813-0} {\bibfield  {journal} {\bibinfo
  {journal} {Nature Physics}\ }\textbf {\bibinfo {volume} {16}},\ \bibinfo
  {pages} {536} (\bibinfo {year} {2020})}\BibitemShut {NoStop}%
\bibitem [{\citenamefont {Zhang}\ \emph {et~al.}(2022)\citenamefont {Zhang},
  \citenamefont {Cui}, \citenamefont {Hutcheon}, \citenamefont {Shipley},
  \citenamefont {Song}, \citenamefont {Du}, \citenamefont {Kresin},
  \citenamefont {Duan}, \citenamefont {Pickard},\ and\ \citenamefont
  {Yao}}]{zhang_design_2022}%
  \BibitemOpen
  \bibfield  {author} {\bibinfo {author} {\bibfnamefont {Z.}~\bibnamefont
  {Zhang}}, \bibinfo {author} {\bibfnamefont {T.}~\bibnamefont {Cui}}, \bibinfo
  {author} {\bibfnamefont {M.~J.}\ \bibnamefont {Hutcheon}}, \bibinfo {author}
  {\bibfnamefont {A.~M.}\ \bibnamefont {Shipley}}, \bibinfo {author}
  {\bibfnamefont {H.}~\bibnamefont {Song}}, \bibinfo {author} {\bibfnamefont
  {M.}~\bibnamefont {Du}}, \bibinfo {author} {\bibfnamefont {V.~Z.}\
  \bibnamefont {Kresin}}, \bibinfo {author} {\bibfnamefont {D.}~\bibnamefont
  {Duan}}, \bibinfo {author} {\bibfnamefont {C.~J.}\ \bibnamefont {Pickard}},\
  and\ \bibinfo {author} {\bibfnamefont {Y.}~\bibnamefont {Yao}},\ }\bibfield
  {title} {\bibinfo {title} {Design {Principles} for {High}-{Temperature}
  {Superconductors} with a {Hydrogen}-{Based} {Alloy} {Backbone} at {Moderate}
  {Pressure}},\ }\href {https://doi.org/10.1103/PhysRevLett.128.047001}
  {\bibfield  {journal} {\bibinfo  {journal} {Physical Review Letters}\
  }\textbf {\bibinfo {volume} {128}},\ \bibinfo {pages} {047001} (\bibinfo
  {year} {2022})}\BibitemShut {NoStop}%
\bibitem [{\citenamefont {Bouchiat}\ \emph {et~al.}(1998)\citenamefont
  {Bouchiat}, \citenamefont {Vion}, \citenamefont {Joyez}, \citenamefont
  {Esteve},\ and\ \citenamefont {Devoret}}]{bouchiat_quantum_1998}%
  \BibitemOpen
  \bibfield  {author} {\bibinfo {author} {\bibfnamefont {V.}~\bibnamefont
  {Bouchiat}}, \bibinfo {author} {\bibfnamefont {D.}~\bibnamefont {Vion}},
  \bibinfo {author} {\bibfnamefont {P.}~\bibnamefont {Joyez}}, \bibinfo
  {author} {\bibfnamefont {D.}~\bibnamefont {Esteve}},\ and\ \bibinfo {author}
  {\bibfnamefont {M.~H.}\ \bibnamefont {Devoret}},\ }\bibfield  {title}
  {\bibinfo {title} {Quantum coherence with a single {Cooper} pair},\ }\href
  {https://doi.org/10.1238/Physica.Topical.076a00165} {\bibfield  {journal}
  {\bibinfo  {journal} {Physica Scripta}\ }\textbf {\bibinfo {volume} {1998}},\
  \bibinfo {pages} {165} (\bibinfo {year} {1998})}\BibitemShut {NoStop}%
\bibitem [{\citenamefont {Nakamura}\ \emph {et~al.}(1999)\citenamefont
  {Nakamura}, \citenamefont {Pashkin},\ and\ \citenamefont
  {Tsai}}]{nakamura_coherent_1999}%
  \BibitemOpen
  \bibfield  {author} {\bibinfo {author} {\bibfnamefont {Y.}~\bibnamefont
  {Nakamura}}, \bibinfo {author} {\bibfnamefont {Y.~A.}\ \bibnamefont
  {Pashkin}},\ and\ \bibinfo {author} {\bibfnamefont {J.~S.}\ \bibnamefont
  {Tsai}},\ }\bibfield  {title} {\bibinfo {title} {Coherent control of
  macroscopic quantum states in a single-{Cooper}-pair box},\ }\href
  {https://doi.org/10.1038/19718} {\bibfield  {journal} {\bibinfo  {journal}
  {Nature}\ }\textbf {\bibinfo {volume} {398}},\ \bibinfo {pages} {786}
  (\bibinfo {year} {1999})}\BibitemShut {NoStop}%
\bibitem [{\citenamefont {Kitaev}(2003)}]{kitaev_fault-tolerant_2003}%
  \BibitemOpen
  \bibfield  {author} {\bibinfo {author} {\bibfnamefont {A.~Y.}\ \bibnamefont
  {Kitaev}},\ }\bibfield  {title} {\bibinfo {title} {Fault-tolerant quantum
  computation by anyons},\ }\href
  {https://doi.org/10.1016/S0003-4916(02)00018-0} {\bibfield  {journal}
  {\bibinfo  {journal} {Annals of Physics}\ }\textbf {\bibinfo {volume}
  {303}},\ \bibinfo {pages} {2} (\bibinfo {year} {2003})}\BibitemShut {NoStop}%
\bibitem [{\citenamefont {Sau}\ \emph {et~al.}(2010)\citenamefont {Sau},
  \citenamefont {Lutchyn}, \citenamefont {Tewari},\ and\ \citenamefont
  {Das~Sarma}}]{sau_generic_2010}%
  \BibitemOpen
  \bibfield  {author} {\bibinfo {author} {\bibfnamefont {J.~D.}\ \bibnamefont
  {Sau}}, \bibinfo {author} {\bibfnamefont {R.~M.}\ \bibnamefont {Lutchyn}},
  \bibinfo {author} {\bibfnamefont {S.}~\bibnamefont {Tewari}},\ and\ \bibinfo
  {author} {\bibfnamefont {S.}~\bibnamefont {Das~Sarma}},\ }\bibfield  {title}
  {\bibinfo {title} {Generic {New} {Platform} for {Topological} {Quantum}
  {Computation} {Using} {Semiconductor} {Heterostructures}},\ }\href
  {https://doi.org/10.1103/PhysRevLett.104.040502} {\bibfield  {journal}
  {\bibinfo  {journal} {Physical Review Letters}\ }\textbf {\bibinfo {volume}
  {104}},\ \bibinfo {pages} {040502} (\bibinfo {year} {2010})}\BibitemShut
  {NoStop}%
\bibitem [{\citenamefont {Dusuel}\ \emph {et~al.}(2011)\citenamefont {Dusuel},
  \citenamefont {Kamfor}, \citenamefont {Orús}, \citenamefont {Schmidt},\ and\
  \citenamefont {Vidal}}]{dusuel_robustness_2011}%
  \BibitemOpen
  \bibfield  {author} {\bibinfo {author} {\bibfnamefont {S.}~\bibnamefont
  {Dusuel}}, \bibinfo {author} {\bibfnamefont {M.}~\bibnamefont {Kamfor}},
  \bibinfo {author} {\bibfnamefont {R.}~\bibnamefont {Orús}}, \bibinfo
  {author} {\bibfnamefont {K.~P.}\ \bibnamefont {Schmidt}},\ and\ \bibinfo
  {author} {\bibfnamefont {J.}~\bibnamefont {Vidal}},\ }\bibfield  {title}
  {\bibinfo {title} {Robustness of a {Perturbed} {Topological} {Phase}},\
  }\href {https://doi.org/10.1103/PhysRevLett.106.107203} {\bibfield  {journal}
  {\bibinfo  {journal} {Physical Review Letters}\ }\textbf {\bibinfo {volume}
  {106}},\ \bibinfo {pages} {107203} (\bibinfo {year} {2011})}\BibitemShut
  {NoStop}%
\bibitem [{\citenamefont {Keldysh}(2017)}]{keldysh_coherent_2017}%
  \BibitemOpen
  \bibfield  {author} {\bibinfo {author} {\bibfnamefont {L.~V.}\ \bibnamefont
  {Keldysh}},\ }\bibfield  {title} {\bibinfo {title} {Coherent states of
  excitons},\ }\href {https://doi.org/10.3367/UFNe.2017.10.038227} {\bibfield
  {journal} {\bibinfo  {journal} {Physics-Uspekhi}\ }\textbf {\bibinfo {volume}
  {60}},\ \bibinfo {pages} {1180} (\bibinfo {year} {2017})}\BibitemShut
  {NoStop}%
\bibitem [{\citenamefont {Fil}\ and\ \citenamefont
  {Shevchenko}(2018)}]{fil_electron-hole_2018}%
  \BibitemOpen
  \bibfield  {author} {\bibinfo {author} {\bibfnamefont {D.~V.}\ \bibnamefont
  {Fil}}\ and\ \bibinfo {author} {\bibfnamefont {S.~I.}\ \bibnamefont
  {Shevchenko}},\ }\bibfield  {title} {\bibinfo {title} {Electron-hole
  {Superconductivity} ({Review})},\ }\href {https://doi.org/10.1063/1.5052674}
  {\bibfield  {journal} {\bibinfo  {journal} {Low Temperature Physics}\
  }\textbf {\bibinfo {volume} {44}},\ \bibinfo {pages} {867} (\bibinfo {year}
  {2018})}\BibitemShut {NoStop}%
\bibitem [{\citenamefont {Schouten}\ \emph {et~al.}(2021)\citenamefont
  {Schouten}, \citenamefont {Sager},\ and\ \citenamefont
  {Mazziotti}}]{schouten_exciton_2021}%
  \BibitemOpen
  \bibfield  {author} {\bibinfo {author} {\bibfnamefont {A.~O.}\ \bibnamefont
  {Schouten}}, \bibinfo {author} {\bibfnamefont {L.~M.}\ \bibnamefont
  {Sager}},\ and\ \bibinfo {author} {\bibfnamefont {D.~A.}\ \bibnamefont
  {Mazziotti}},\ }\bibfield  {title} {\bibinfo {title} {Exciton {Condensation}
  in {Molecular}-{Scale} van der {Waals} {Stacks}},\ }\href
  {https://doi.org/10.1021/acs.jpclett.1c02368} {\bibfield  {journal} {\bibinfo
   {journal} {The Journal of Physical Chemistry Letters}\ }\textbf {\bibinfo
  {volume} {12}},\ \bibinfo {pages} {9906} (\bibinfo {year}
  {2021})}\BibitemShut {NoStop}%
\bibitem [{\citenamefont {Muraviev}\ and\ \citenamefont
  {Rumyantsev}(2022)}]{muraviev_exciton_2022}%
  \BibitemOpen
  \bibfield  {author} {\bibinfo {author} {\bibfnamefont {S.~E.}\ \bibnamefont
  {Muraviev}}\ and\ \bibinfo {author} {\bibfnamefont {O.~A.}\ \bibnamefont
  {Rumyantsev}},\ }\bibfield  {title} {\bibinfo {title} {Exciton and
  fermion-pair condensation at non-zero temperature},\ }\href
  {https://doi.org/10.1088/1361-648X/ac4c63} {\bibfield  {journal} {\bibinfo
  {journal} {Journal of Physics: Condensed Matter}\ }\textbf {\bibinfo {volume}
  {34}},\ \bibinfo {pages} {145402} (\bibinfo {year} {2022})}\BibitemShut
  {NoStop}%
\bibitem [{\citenamefont {Sachdev}(2011)}]{sachdev_quantum_2011}%
  \BibitemOpen
  \bibfield  {author} {\bibinfo {author} {\bibfnamefont {S.}~\bibnamefont
  {Sachdev}},\ }\href {www.cambridge.org/9780521514682} {\emph {\bibinfo
  {title} {Quantum {Phase} {Transitions}}}},\ \bibinfo {edition} {2nd}\ ed.\
  (\bibinfo  {publisher} {Cambridge University Press},\ \bibinfo {address}
  {Cambridge, UK},\ \bibinfo {year} {2011})\BibitemShut {NoStop}%
\bibitem [{\citenamefont {Sager}\ and\ \citenamefont
  {Mazziotti}(2022{\natexlab{a}})}]{sager_simultaneous_2022}%
  \BibitemOpen
  \bibfield  {author} {\bibinfo {author} {\bibfnamefont {L.~M.}\ \bibnamefont
  {Sager}}\ and\ \bibinfo {author} {\bibfnamefont {D.~A.}\ \bibnamefont
  {Mazziotti}},\ }\bibfield  {title} {\bibinfo {title} {Simultaneous fermion
  and exciton condensations from a model {Hamiltonian}},\ }\href
  {https://doi.org/10.1103/PhysRevB.105.035143} {\bibfield  {journal} {\bibinfo
   {journal} {Physical Review B}\ }\textbf {\bibinfo {volume} {105}},\ \bibinfo
  {pages} {035143} (\bibinfo {year} {2022}{\natexlab{a}})}\BibitemShut
  {NoStop}%
\bibitem [{\citenamefont {Bardeen}\ \emph {et~al.}(1957)\citenamefont
  {Bardeen}, \citenamefont {Cooper},\ and\ \citenamefont
  {Schrieffer}}]{bardeen_theory_1957}%
  \BibitemOpen
  \bibfield  {author} {\bibinfo {author} {\bibfnamefont {J.}~\bibnamefont
  {Bardeen}}, \bibinfo {author} {\bibfnamefont {L.~N.}\ \bibnamefont
  {Cooper}},\ and\ \bibinfo {author} {\bibfnamefont {J.~R.}\ \bibnamefont
  {Schrieffer}},\ }\bibfield  {title} {\bibinfo {title} {Theory of
  {Superconductivity}},\ }\href {https://doi.org/10.1103/PhysRev.108.1175}
  {\bibfield  {journal} {\bibinfo  {journal} {Physical Review}\ }\textbf
  {\bibinfo {volume} {108}},\ \bibinfo {pages} {1175} (\bibinfo {year}
  {1957})}\BibitemShut {NoStop}%
\bibitem [{\citenamefont {Lipkin}\ \emph {et~al.}(1965)\citenamefont {Lipkin},
  \citenamefont {Meshkov},\ and\ \citenamefont {Glick}}]{lipkin_validity_1965}%
  \BibitemOpen
  \bibfield  {author} {\bibinfo {author} {\bibfnamefont {H.~J.}\ \bibnamefont
  {Lipkin}}, \bibinfo {author} {\bibfnamefont {N.}~\bibnamefont {Meshkov}},\
  and\ \bibinfo {author} {\bibfnamefont {A.~J.}\ \bibnamefont {Glick}},\
  }\bibfield  {title} {\bibinfo {title} {Validity of many-body approximation
  methods for a solvable model: ({I}). {Exact} solutions and perturbation
  theory},\ }\href {https://doi.org/10.1016/0029-5582(65)90862-X} {\bibfield
  {journal} {\bibinfo  {journal} {Nuclear Physics}\ }\textbf {\bibinfo {volume}
  {62}},\ \bibinfo {pages} {188} (\bibinfo {year} {1965})}\BibitemShut
  {NoStop}%
\bibitem [{\citenamefont {Meshkov}\ \emph {et~al.}(1965)\citenamefont
  {Meshkov}, \citenamefont {Glick},\ and\ \citenamefont
  {Lipkin}}]{meshkov_validity_1965}%
  \BibitemOpen
  \bibfield  {author} {\bibinfo {author} {\bibfnamefont {N.}~\bibnamefont
  {Meshkov}}, \bibinfo {author} {\bibfnamefont {A.~J.}\ \bibnamefont {Glick}},\
  and\ \bibinfo {author} {\bibfnamefont {H.~J.}\ \bibnamefont {Lipkin}},\
  }\bibfield  {title} {\bibinfo {title} {Validity of many-body approximation
  methods for a solvable model: ({II}). {Linearization} procedures},\ }\href
  {https://doi.org/10.1016/0029-5582(65)90863-1} {\bibfield  {journal}
  {\bibinfo  {journal} {Nuclear Physics}\ }\textbf {\bibinfo {volume} {62}},\
  \bibinfo {pages} {199} (\bibinfo {year} {1965})}\BibitemShut {NoStop}%
\bibitem [{\citenamefont {Glick}\ \emph {et~al.}(1965)\citenamefont {Glick},
  \citenamefont {Lipkin},\ and\ \citenamefont {Meshkov}}]{glick_validity_1965}%
  \BibitemOpen
  \bibfield  {author} {\bibinfo {author} {\bibfnamefont {A.~J.}\ \bibnamefont
  {Glick}}, \bibinfo {author} {\bibfnamefont {H.~J.}\ \bibnamefont {Lipkin}},\
  and\ \bibinfo {author} {\bibfnamefont {N.}~\bibnamefont {Meshkov}},\
  }\bibfield  {title} {\bibinfo {title} {Validity of many-body approximation
  methods for a solvable model: ({III}). {Diagram} summations},\ }\href
  {https://doi.org/10.1016/0029-5582(65)90864-3} {\bibfield  {journal}
  {\bibinfo  {journal} {Nuclear Physics}\ }\textbf {\bibinfo {volume} {62}},\
  \bibinfo {pages} {211} (\bibinfo {year} {1965})}\BibitemShut {NoStop}%
\bibitem [{\citenamefont
  {Mazziotti}(2004{\natexlab{a}})}]{mazziotti_exactness_2004}%
  \BibitemOpen
  \bibfield  {author} {\bibinfo {author} {\bibfnamefont {D.~A.}\ \bibnamefont
  {Mazziotti}},\ }\bibfield  {title} {\bibinfo {title} {Exactness of wave
  functions from two-body exponential transformations in many-body quantum
  theory},\ }\href {https://doi.org/10.1103/PhysRevA.69.012507} {\bibfield
  {journal} {\bibinfo  {journal} {Physical Review A}\ }\textbf {\bibinfo
  {volume} {69}},\ \bibinfo {pages} {012507} (\bibinfo {year}
  {2004}{\natexlab{a}})}\BibitemShut {NoStop}%
\bibitem [{\citenamefont {Heiss}\ \emph {et~al.}(2005)\citenamefont {Heiss},
  \citenamefont {Scholtz},\ and\ \citenamefont {Geyer}}]{heiss_large_2005}%
  \BibitemOpen
  \bibfield  {author} {\bibinfo {author} {\bibfnamefont {W.~D.}\ \bibnamefont
  {Heiss}}, \bibinfo {author} {\bibfnamefont {F.~G.}\ \bibnamefont {Scholtz}},\
  and\ \bibinfo {author} {\bibfnamefont {H.~B.}\ \bibnamefont {Geyer}},\
  }\bibfield  {title} {\bibinfo {title} {The large {N} behaviour of the
  {Lipkin} model and exceptional points},\ }\href
  {https://doi.org/10.1088/0305-4470/38/9/002} {\bibfield  {journal} {\bibinfo
  {journal} {Journal of Physics A: Mathematical and General}\ }\textbf
  {\bibinfo {volume} {38}},\ \bibinfo {pages} {1843} (\bibinfo {year}
  {2005})}\BibitemShut {NoStop}%
\bibitem [{\citenamefont {Castaños}\ \emph {et~al.}(2006)\citenamefont
  {Castaños}, \citenamefont {López-Peña}, \citenamefont {Hirsch},\ and\
  \citenamefont {López-Moreno}}]{castanos_classical_2006}%
  \BibitemOpen
  \bibfield  {author} {\bibinfo {author} {\bibfnamefont {O.}~\bibnamefont
  {Castaños}}, \bibinfo {author} {\bibfnamefont {R.}~\bibnamefont
  {López-Peña}}, \bibinfo {author} {\bibfnamefont {J.~G.}\ \bibnamefont
  {Hirsch}},\ and\ \bibinfo {author} {\bibfnamefont {E.}~\bibnamefont
  {López-Moreno}},\ }\bibfield  {title} {\bibinfo {title} {Classical and
  quantum phase transitions in the {Lipkin}-{Meshkov}-{Glick} model},\ }\href
  {https://doi.org/10.1103/PhysRevB.74.104118} {\bibfield  {journal} {\bibinfo
  {journal} {Physical Review B}\ }\textbf {\bibinfo {volume} {74}},\ \bibinfo
  {pages} {104118} (\bibinfo {year} {2006})}\BibitemShut {NoStop}%
\bibitem [{\citenamefont {Gidofalvi}\ and\ \citenamefont
  {Mazziotti}(2006)}]{gidofalvi_computation_2006}%
  \BibitemOpen
  \bibfield  {author} {\bibinfo {author} {\bibfnamefont {G.}~\bibnamefont
  {Gidofalvi}}\ and\ \bibinfo {author} {\bibfnamefont {D.~A.}\ \bibnamefont
  {Mazziotti}},\ }\bibfield  {title} {\bibinfo {title} {Computation of quantum
  phase transitions by reduced-density-matrix mechanics},\ }\href
  {https://doi.org/10.1103/PhysRevA.74.012501} {\bibfield  {journal} {\bibinfo
  {journal} {Physical Review A}\ }\textbf {\bibinfo {volume} {74}},\ \bibinfo
  {pages} {012501} (\bibinfo {year} {2006})}\BibitemShut {NoStop}%
\bibitem [{\citenamefont {Chen}\ and\ \citenamefont
  {Liang}(2006)}]{chen_unconventional_2006}%
  \BibitemOpen
  \bibfield  {author} {\bibinfo {author} {\bibfnamefont {G.}~\bibnamefont
  {Chen}}\ and\ \bibinfo {author} {\bibfnamefont {J.-Q.}\ \bibnamefont
  {Liang}},\ }\bibfield  {title} {\bibinfo {title} {Unconventional quantum
  phase transition in the finite-size {Lipkin}–{Meshkov}–{Glick} model},\
  }\href {https://doi.org/10.1088/1367-2630/8/12/297} {\bibfield  {journal}
  {\bibinfo  {journal} {New Journal of Physics}\ }\textbf {\bibinfo {volume}
  {8}},\ \bibinfo {pages} {297} (\bibinfo {year} {2006})}\BibitemShut {NoStop}%
\bibitem [{\citenamefont {Romera}\ \emph {et~al.}(2014)\citenamefont {Romera},
  \citenamefont {Calixto},\ and\ \citenamefont
  {Castaños}}]{romera_phase_2014}%
  \BibitemOpen
  \bibfield  {author} {\bibinfo {author} {\bibfnamefont {E.}~\bibnamefont
  {Romera}}, \bibinfo {author} {\bibfnamefont {M.}~\bibnamefont {Calixto}},\
  and\ \bibinfo {author} {\bibfnamefont {O.}~\bibnamefont {Castaños}},\
  }\bibfield  {title} {\bibinfo {title} {Phase space analysis of first-,
  second- and third-order quantum phase transitions in the
  {Lipkin}–{Meshkov}–{Glick} model},\ }\href
  {https://doi.org/10.1088/0031-8949/89/9/095103} {\bibfield  {journal}
  {\bibinfo  {journal} {Physica Scripta}\ }\textbf {\bibinfo {volume} {89}},\
  \bibinfo {pages} {095103} (\bibinfo {year} {2014})},\ \bibinfo {note}
  {publisher: IOP Publishing}\BibitemShut {NoStop}%
\bibitem [{\citenamefont {Wang}\ and\ \citenamefont
  {Pérez-Bernal}(2021)}]{wang_characterizing_2021}%
  \BibitemOpen
  \bibfield  {author} {\bibinfo {author} {\bibfnamefont {Q.}~\bibnamefont
  {Wang}}\ and\ \bibinfo {author} {\bibfnamefont {F.}~\bibnamefont
  {Pérez-Bernal}},\ }\bibfield  {title} {\bibinfo {title} {Characterizing the
  {Lipkin}-{Meshkov}-{Glick} model excited-state quantum phase transition using
  dynamical and statistical properties of the diagonal entropy},\ }\href
  {https://doi.org/10.1103/PhysRevE.103.032109} {\bibfield  {journal} {\bibinfo
   {journal} {Physical Review E}\ }\textbf {\bibinfo {volume} {103}},\ \bibinfo
  {pages} {032109} (\bibinfo {year} {2021})}\BibitemShut {NoStop}%
\bibitem [{\citenamefont {Sager}\ \emph
  {et~al.}(2020{\natexlab{a}})\citenamefont {Sager}, \citenamefont {Smart},\
  and\ \citenamefont {Mazziotti}}]{sager_preparation_2020}%
  \BibitemOpen
  \bibfield  {author} {\bibinfo {author} {\bibfnamefont {L.~M.}\ \bibnamefont
  {Sager}}, \bibinfo {author} {\bibfnamefont {S.~E.}\ \bibnamefont {Smart}},\
  and\ \bibinfo {author} {\bibfnamefont {D.~A.}\ \bibnamefont {Mazziotti}},\
  }\bibfield  {title} {\bibinfo {title} {Preparation of an {Exciton}
  {Condensate} of {Photons} on a 53-{Qubit} {Quantum} {Computer}},\ }\href
  {https://doi.org/10.1103/PhysRevResearch.2.043205} {\bibfield  {journal}
  {\bibinfo  {journal} {Physical Review Research}\ }\textbf {\bibinfo {volume}
  {2}},\ \bibinfo {pages} {043205} (\bibinfo {year}
  {2020}{\natexlab{a}})}\BibitemShut {NoStop}%
\bibitem [{\citenamefont {Sager}\ \emph
  {et~al.}(2020{\natexlab{b}})\citenamefont {Sager}, \citenamefont {Safaei},\
  and\ \citenamefont {Mazziotti}}]{sager_potential_2020}%
  \BibitemOpen
  \bibfield  {author} {\bibinfo {author} {\bibfnamefont {L.~M.}\ \bibnamefont
  {Sager}}, \bibinfo {author} {\bibfnamefont {S.}~\bibnamefont {Safaei}},\ and\
  \bibinfo {author} {\bibfnamefont {D.~A.}\ \bibnamefont {Mazziotti}},\
  }\bibfield  {title} {\bibinfo {title} {Potential coexistence of exciton and
  fermion-pair condensations},\ }\href
  {https://doi.org/10.1103/PhysRevB.101.081107} {\bibfield  {journal} {\bibinfo
   {journal} {Physical Review B}\ }\textbf {\bibinfo {volume} {101}},\ \bibinfo
  {pages} {081107} (\bibinfo {year} {2020}{\natexlab{b}})}\BibitemShut
  {NoStop}%
\bibitem [{\citenamefont {Wu}\ \emph {et~al.}(2004)\citenamefont {Wu},
  \citenamefont {Sarandy},\ and\ \citenamefont {Lidar}}]{wu_quantum_2004}%
  \BibitemOpen
  \bibfield  {author} {\bibinfo {author} {\bibfnamefont {L.-A.}\ \bibnamefont
  {Wu}}, \bibinfo {author} {\bibfnamefont {M.~S.}\ \bibnamefont {Sarandy}},\
  and\ \bibinfo {author} {\bibfnamefont {D.~A.}\ \bibnamefont {Lidar}},\
  }\bibfield  {title} {\bibinfo {title} {Quantum {Phase} {Transitions} and
  {Bipartite} {Entanglement}},\ }\href
  {https://doi.org/10.1103/PhysRevLett.93.250404} {\bibfield  {journal}
  {\bibinfo  {journal} {Physical Review Letters}\ }\textbf {\bibinfo {volume}
  {93}},\ \bibinfo {pages} {250404} (\bibinfo {year} {2004})},\ \bibinfo {note}
  {publisher: American Physical Society}\BibitemShut {NoStop}%
\bibitem [{\citenamefont {Schwerdtfeger}\ and\ \citenamefont
  {Mazziotti}(2009)}]{schwerdtfeger_convex-set_2009}%
  \BibitemOpen
  \bibfield  {author} {\bibinfo {author} {\bibfnamefont {C.~A.}\ \bibnamefont
  {Schwerdtfeger}}\ and\ \bibinfo {author} {\bibfnamefont {D.~A.}\ \bibnamefont
  {Mazziotti}},\ }\bibfield  {title} {\bibinfo {title} {Convex-set description
  of quantum phase transitions in the transverse {Ising} model using
  reduced-density-matrix theory},\ }\href {https://doi.org/10.1063/1.3143403}
  {\bibfield  {journal} {\bibinfo  {journal} {The Journal of Chemical Physics}\
  }\textbf {\bibinfo {volume} {130}},\ \bibinfo {pages} {224102} (\bibinfo
  {year} {2009})}\BibitemShut {NoStop}%
\bibitem [{\citenamefont {Zauner}\ \emph {et~al.}(2016)\citenamefont {Zauner},
  \citenamefont {Draxler}, \citenamefont {Vanderstraeten}, \citenamefont
  {Haegeman},\ and\ \citenamefont {Verstraete}}]{zauner_symmetry_2016}%
  \BibitemOpen
  \bibfield  {author} {\bibinfo {author} {\bibfnamefont {V.}~\bibnamefont
  {Zauner}}, \bibinfo {author} {\bibfnamefont {D.}~\bibnamefont {Draxler}},
  \bibinfo {author} {\bibfnamefont {L.}~\bibnamefont {Vanderstraeten}},
  \bibinfo {author} {\bibfnamefont {J.}~\bibnamefont {Haegeman}},\ and\
  \bibinfo {author} {\bibfnamefont {F.}~\bibnamefont {Verstraete}},\ }\bibfield
   {title} {\bibinfo {title} {Symmetry breaking and the geometry of reduced
  density matrices},\ }\href {https://doi.org/10.1088/1367-2630/18/11/113033}
  {\bibfield  {journal} {\bibinfo  {journal} {New Journal of Physics}\ }\textbf
  {\bibinfo {volume} {18}},\ \bibinfo {pages} {113033} (\bibinfo {year}
  {2016})}\BibitemShut {NoStop}%
\bibitem [{\citenamefont {Yang}(1962)}]{yang_concept_1962}%
  \BibitemOpen
  \bibfield  {author} {\bibinfo {author} {\bibfnamefont {C.~N.}\ \bibnamefont
  {Yang}},\ }\bibfield  {title} {\bibinfo {title} {Concept of {Off}-{Diagonal}
  {Long}-{Range} {Order} and the {Quantum} {Phases} of {Liquid} {He} and of
  {Superconductors}},\ }\href {https://doi.org/10.1103/RevModPhys.34.694}
  {\bibfield  {journal} {\bibinfo  {journal} {Reviews of Modern Physics}\
  }\textbf {\bibinfo {volume} {34}},\ \bibinfo {pages} {694} (\bibinfo {year}
  {1962})}\BibitemShut {NoStop}%
\bibitem [{\citenamefont {Sasaki}(1965)}]{sasaki_eigenvalues_1965}%
  \BibitemOpen
  \bibfield  {author} {\bibinfo {author} {\bibfnamefont {F.}~\bibnamefont
  {Sasaki}},\ }\bibfield  {title} {\bibinfo {title} {Eigenvalues of {Fermion}
  {Density} {Matrices}},\ }\href {https://doi.org/10.1103/PhysRev.138.B1338}
  {\bibfield  {journal} {\bibinfo  {journal} {Physical Review}\ }\textbf
  {\bibinfo {volume} {138}},\ \bibinfo {pages} {B1338} (\bibinfo {year}
  {1965})}\BibitemShut {NoStop}%
\bibitem [{\citenamefont {Coleman}(1965)}]{coleman_structure_1965}%
  \BibitemOpen
  \bibfield  {author} {\bibinfo {author} {\bibfnamefont {A.~J.}\ \bibnamefont
  {Coleman}},\ }\bibfield  {title} {\bibinfo {title} {Structure of {Fermion}
  {Density} {Matrices}. {II}. {Antisymmetrized} {Geminal} {Powers}},\ }\href
  {https://doi.org/10.1063/1.1704794} {\bibfield  {journal} {\bibinfo
  {journal} {Journal of Mathematical Physics}\ }\textbf {\bibinfo {volume}
  {6}},\ \bibinfo {pages} {1425} (\bibinfo {year} {1965})}\BibitemShut
  {NoStop}%
\bibitem [{\citenamefont {Garrod}\ and\ \citenamefont
  {Rosina}(1969)}]{garrod_particlehole_1969}%
  \BibitemOpen
  \bibfield  {author} {\bibinfo {author} {\bibfnamefont {C.}~\bibnamefont
  {Garrod}}\ and\ \bibinfo {author} {\bibfnamefont {M.}~\bibnamefont
  {Rosina}},\ }\bibfield  {title} {\bibinfo {title} {Particle‐{Hole}
  {Matrix}: {Its} {Connection} with the {Symmetries} and {Collective}
  {Features} of the {Ground} {State}},\ }\href
  {https://doi.org/10.1063/1.1664770} {\bibfield  {journal} {\bibinfo
  {journal} {Journal of Mathematical Physics}\ }\textbf {\bibinfo {volume}
  {10}},\ \bibinfo {pages} {1855} (\bibinfo {year} {1969})}\BibitemShut
  {NoStop}%
\bibitem [{\citenamefont {Safaei}\ and\ \citenamefont
  {Mazziotti}(2018)}]{safaei_quantum_2018}%
  \BibitemOpen
  \bibfield  {author} {\bibinfo {author} {\bibfnamefont {S.}~\bibnamefont
  {Safaei}}\ and\ \bibinfo {author} {\bibfnamefont {D.~A.}\ \bibnamefont
  {Mazziotti}},\ }\bibfield  {title} {\bibinfo {title} {Quantum signature of
  exciton condensation},\ }\href {https://doi.org/10.1103/PhysRevB.98.045122}
  {\bibfield  {journal} {\bibinfo  {journal} {Physical Review B}\ }\textbf
  {\bibinfo {volume} {98}},\ \bibinfo {pages} {045122} (\bibinfo {year}
  {2018})}\BibitemShut {NoStop}%
\bibitem [{\citenamefont {Erdahl}\ and\ \citenamefont
  {Jin}(2000)}]{erdahl_lower_2000}%
  \BibitemOpen
  \bibfield  {author} {\bibinfo {author} {\bibfnamefont {R.~M.}\ \bibnamefont
  {Erdahl}}\ and\ \bibinfo {author} {\bibfnamefont {B.}~\bibnamefont {Jin}},\
  }\bibfield  {title} {\bibinfo {title} {The lower bound method for reduced
  density matrices},\ }\href {https://doi.org/10.1016/S0166-1280(00)00494-2}
  {\bibfield  {journal} {\bibinfo  {journal} {Journal of Molecular Structure:
  THEOCHEM}\ }\textbf {\bibinfo {volume} {527}},\ \bibinfo {pages} {207}
  (\bibinfo {year} {2000})}\BibitemShut {NoStop}%
\bibitem [{\citenamefont {Warren}\ \emph {et~al.}(2022)\citenamefont {Warren},
  \citenamefont {Sager-Smith},\ and\ \citenamefont {Mazziotti}}]{Warren2022}%
  \BibitemOpen
  \bibfield  {author} {\bibinfo {author} {\bibfnamefont {S.}~\bibnamefont
  {Warren}}, \bibinfo {author} {\bibfnamefont {L.~M.}\ \bibnamefont
  {Sager-Smith}},\ and\ \bibinfo {author} {\bibfnamefont {D.~A.}\ \bibnamefont
  {Mazziotti}},\ }\bibfield  {title} {\bibinfo {title} {Quantum simulation of
  quantum phase transitions using the convex geometry of reduced density
  matrices},\ }\href {https://doi.org/10.1103/PhysRevA.106.012434} {\bibfield
  {journal} {\bibinfo  {journal} {Phys. Rev. A}\ }\textbf {\bibinfo {volume}
  {106}},\ \bibinfo {pages} {012434} (\bibinfo {year} {2022})}\BibitemShut
  {NoStop}%
\bibitem [{\citenamefont {Penrose}\ and\ \citenamefont
  {Onsager}(1956)}]{penrose_bose-einstein_1956}%
  \BibitemOpen
  \bibfield  {author} {\bibinfo {author} {\bibfnamefont {O.}~\bibnamefont
  {Penrose}}\ and\ \bibinfo {author} {\bibfnamefont {L.}~\bibnamefont
  {Onsager}},\ }\bibfield  {title} {\bibinfo {title} {Bose-{Einstein}
  {Condensation} and {Liquid} {Helium}},\ }\href
  {https://doi.org/10.1103/PhysRev.104.576} {\bibfield  {journal} {\bibinfo
  {journal} {Physical Review}\ }\textbf {\bibinfo {volume} {104}},\ \bibinfo
  {pages} {576} (\bibinfo {year} {1956})},\ \bibinfo {note} {publisher:
  American Physical Society}\BibitemShut {NoStop}%
\bibitem [{\citenamefont {{Bose}}(1924)}]{bose_plancks_1924}%
  \BibitemOpen
  \bibfield  {author} {\bibinfo {author} {\bibnamefont {{Bose}}},\ }\bibfield
  {title} {\bibinfo {title} {Plancks {Gesetz} und {Lichtquantenhypothese}},\
  }\href {https://doi.org/10.1007/BF01327326} {\bibfield  {journal} {\bibinfo
  {journal} {Zeitschrift für Physik}\ }\textbf {\bibinfo {volume} {26}},\
  \bibinfo {pages} {178} (\bibinfo {year} {1924})}\BibitemShut {NoStop}%
\bibitem [{\citenamefont {Surjan}(1999)}]{surjan_introduction_1999}%
  \BibitemOpen
  \bibfield  {author} {\bibinfo {author} {\bibfnamefont {P.~R.}\ \bibnamefont
  {Surjan}},\ }\bibfield  {title} {\bibinfo {title} {An {Introduction} to the
  {Theory} of {Geminals}},\ }in\ \href@noop {} {\emph {\bibinfo {booktitle}
  {Correlation and {Localization}}}}\ (\bibinfo  {publisher} {Springer},\
  \bibinfo {address} {Berlin, Heidelberg},\ \bibinfo {year} {1999})\ pp.\
  \bibinfo {pages} {63--88}\BibitemShut {NoStop}%
\bibitem [{\citenamefont {London}(1938)}]{london_bose-einstein_1938}%
  \BibitemOpen
  \bibfield  {author} {\bibinfo {author} {\bibfnamefont {F.}~\bibnamefont
  {London}},\ }\bibfield  {title} {\bibinfo {title} {On the {Bose}-{Einstein}
  {Condensation}},\ }\href {https://doi.org/10.1103/PhysRev.54.947} {\bibfield
  {journal} {\bibinfo  {journal} {Physical Review}\ }\textbf {\bibinfo {volume}
  {54}},\ \bibinfo {pages} {947} (\bibinfo {year} {1938})}\BibitemShut
  {NoStop}%
\bibitem [{\citenamefont {Tisza}(1947)}]{tisza_theory_1947}%
  \BibitemOpen
  \bibfield  {author} {\bibinfo {author} {\bibfnamefont {L.}~\bibnamefont
  {Tisza}},\ }\bibfield  {title} {\bibinfo {title} {The {Theory} of {Liquid}
  {Helium}},\ }\href {https://doi.org/10.1103/PhysRev.72.838} {\bibfield
  {journal} {\bibinfo  {journal} {Physical Review}\ }\textbf {\bibinfo {volume}
  {72}},\ \bibinfo {pages} {838} (\bibinfo {year} {1947})}\BibitemShut
  {NoStop}%
\bibitem [{\citenamefont {Raeber}\ and\ \citenamefont
  {Mazziotti}(2015)}]{raeber_large_2015}%
  \BibitemOpen
  \bibfield  {author} {\bibinfo {author} {\bibfnamefont {A.}~\bibnamefont
  {Raeber}}\ and\ \bibinfo {author} {\bibfnamefont {D.~A.}\ \bibnamefont
  {Mazziotti}},\ }\bibfield  {title} {\bibinfo {title} {Large eigenvalue of the
  cumulant part of the two-electron reduced density matrix as a measure of
  off-diagonal long-range order},\ }\href
  {https://doi.org/10.1103/PhysRevA.92.052502} {\bibfield  {journal} {\bibinfo
  {journal} {Physical Review A}\ }\textbf {\bibinfo {volume} {92}},\ \bibinfo
  {pages} {052502} (\bibinfo {year} {2015})}\BibitemShut {NoStop}%
\bibitem [{\citenamefont {Sager}\ and\ \citenamefont
  {Mazziotti}(2021)}]{sager_superconductivity_2021}%
  \BibitemOpen
  \bibfield  {author} {\bibinfo {author} {\bibfnamefont {L.~M.}\ \bibnamefont
  {Sager}}\ and\ \bibinfo {author} {\bibfnamefont {D.~A.}\ \bibnamefont
  {Mazziotti}},\ }\bibfield  {title} {\bibinfo {title} {Superconductivity and
  {Non}-{Classical} {Long}-{Range} {Order} on a {Quantum} {Computer}},\ }\href
  {http://arxiv.org/abs/2102.08960} {\bibfield  {journal} {\bibinfo  {journal}
  {arXiv:2102.08960 [physics, physics:quant-ph]}\ } (\bibinfo {year}
  {2021})}\BibitemShut {NoStop}%
\bibitem [{\citenamefont {Sager}\ and\ \citenamefont
  {Mazziotti}(2022{\natexlab{b}})}]{sager_cooper-pair_2022}%
  \BibitemOpen
  \bibfield  {author} {\bibinfo {author} {\bibfnamefont {L.~M.}\ \bibnamefont
  {Sager}}\ and\ \bibinfo {author} {\bibfnamefont {D.~A.}\ \bibnamefont
  {Mazziotti}},\ }\bibfield  {title} {\bibinfo {title} {Cooper-pair condensates
  with nonclassical long-range order on quantum devices},\ }\href
  {https://doi.org/10.1103/PhysRevResearch.4.013003} {\bibfield  {journal}
  {\bibinfo  {journal} {Physical Review Research}\ }\textbf {\bibinfo {volume}
  {4}},\ \bibinfo {pages} {013003} (\bibinfo {year}
  {2022}{\natexlab{b}})}\BibitemShut {NoStop}%
\bibitem [{\citenamefont {Sager}\ \emph {et~al.}(2022)\citenamefont {Sager},
  \citenamefont {Schouten},\ and\ \citenamefont
  {Mazziotti}}]{sager_beginnings_2022}%
  \BibitemOpen
  \bibfield  {author} {\bibinfo {author} {\bibfnamefont {L.~M.}\ \bibnamefont
  {Sager}}, \bibinfo {author} {\bibfnamefont {A.~O.}\ \bibnamefont
  {Schouten}},\ and\ \bibinfo {author} {\bibfnamefont {D.~A.}\ \bibnamefont
  {Mazziotti}},\ }\bibfield  {title} {\bibinfo {title} {Beginnings of exciton
  condensation in coronene analog of graphene double layer},\ }\href
  {https://doi.org/10.1063/5.0084564} {\bibfield  {journal} {\bibinfo
  {journal} {The Journal of Chemical Physics}\ }\textbf {\bibinfo {volume}
  {156}},\ \bibinfo {pages} {154702} (\bibinfo {year} {2022})},\ \bibinfo
  {note} {publisher: American Institute of Physics}\BibitemShut {NoStop}%
\bibitem [{\citenamefont
  {Mazziotti}(2006{\natexlab{a}})}]{mazziotti_quantum_2006}%
  \BibitemOpen
  \bibfield  {author} {\bibinfo {author} {\bibfnamefont {D.~A.}\ \bibnamefont
  {Mazziotti}},\ }\bibfield  {title} {\bibinfo {title} {Quantum {Chemisty}
  without {Wavefunctions}: {Two}-{Electron} {Reduced} {Density} {Matrices}},\
  }\href {https://doi.org/10.1021/ar050029d} {\bibfield  {journal} {\bibinfo
  {journal} {Accounts of Chemical Research}\ }\textbf {\bibinfo {volume}
  {39}},\ \bibinfo {pages} {207} (\bibinfo {year}
  {2006}{\natexlab{a}})}\BibitemShut {NoStop}%
\bibitem [{\citenamefont
  {Mazziotti}(2004{\natexlab{b}})}]{mazziotti_realization_2004}%
  \BibitemOpen
  \bibfield  {author} {\bibinfo {author} {\bibfnamefont {D.~A.}\ \bibnamefont
  {Mazziotti}},\ }\bibfield  {title} {\bibinfo {title} {Realization of
  {Quantum} {Chemistry} without {Wave} {Functions} through {First}-{Order}
  {Semidefinite} {Programming}},\ }\href
  {https://doi.org/10.1103/PhysRevLett.93.213001} {\bibfield  {journal}
  {\bibinfo  {journal} {Physical Review Letters}\ }\textbf {\bibinfo {volume}
  {93}},\ \bibinfo {pages} {213001} (\bibinfo {year}
  {2004}{\natexlab{b}})}\BibitemShut {NoStop}%
\bibitem [{\citenamefont
  {Mazziotti}(2006{\natexlab{b}})}]{mazziotti_anti-hermitian_2006}%
  \BibitemOpen
  \bibfield  {author} {\bibinfo {author} {\bibfnamefont {D.~A.}\ \bibnamefont
  {Mazziotti}},\ }\bibfield  {title} {\bibinfo {title} {Anti-{Hermitian}
  {Contracted} {Schr}{\"o}dinger {Equation}: {Direct} {Determination} of the
  {Two}-{Electron} {Reduced} {Density} {Matrices} of {Many}-{Electron}
  {Molecules}},\ }\href {https://doi.org/10.1103/PhysRevLett.97.143002}
  {\bibfield  {journal} {\bibinfo  {journal} {Physical Review Letters}\
  }\textbf {\bibinfo {volume} {97}},\ \bibinfo {pages} {143002} (\bibinfo
  {year} {2006}{\natexlab{b}})}\BibitemShut {NoStop}%
\bibitem [{\citenamefont {Mazziotti}(2011)}]{mazziotti_large-scale_2011}%
  \BibitemOpen
  \bibfield  {author} {\bibinfo {author} {\bibfnamefont {D.~A.}\ \bibnamefont
  {Mazziotti}},\ }\bibfield  {title} {\bibinfo {title} {Large-{Scale}
  {Semidefinite} {Programming} for {Many}-{Electron} {Quantum} {Mechanics}},\
  }\href {https://doi.org/10.1103/PhysRevLett.106.083001} {\bibfield  {journal}
  {\bibinfo  {journal} {Physical Review Letters}\ }\textbf {\bibinfo {volume}
  {106}},\ \bibinfo {pages} {083001} (\bibinfo {year} {2011})}\BibitemShut
  {NoStop}%
\bibitem [{\citenamefont {Coleman}(1963)}]{coleman_structure_1963}%
  \BibitemOpen
  \bibfield  {author} {\bibinfo {author} {\bibfnamefont {A.~J.}\ \bibnamefont
  {Coleman}},\ }\bibfield  {title} {\bibinfo {title} {Structure of {Fermion}
  {Density} {Matrices}},\ }\href {https://doi.org/10.1103/RevModPhys.35.668}
  {\bibfield  {journal} {\bibinfo  {journal} {Reviews of Modern Physics}\
  }\textbf {\bibinfo {volume} {35}},\ \bibinfo {pages} {668} (\bibinfo {year}
  {1963})}\BibitemShut {NoStop}%
\bibitem [{\citenamefont {Chen}\ \emph {et~al.}(2016)\citenamefont {Chen},
  \citenamefont {Ji}, \citenamefont {Liu}, \citenamefont {Shen},\ and\
  \citenamefont {Zeng}}]{chen_geometry_2016}%
  \BibitemOpen
  \bibfield  {author} {\bibinfo {author} {\bibfnamefont {J.-Y.}\ \bibnamefont
  {Chen}}, \bibinfo {author} {\bibfnamefont {Z.}~\bibnamefont {Ji}}, \bibinfo
  {author} {\bibfnamefont {Z.-X.}\ \bibnamefont {Liu}}, \bibinfo {author}
  {\bibfnamefont {Y.}~\bibnamefont {Shen}},\ and\ \bibinfo {author}
  {\bibfnamefont {B.}~\bibnamefont {Zeng}},\ }\bibfield  {title} {\bibinfo
  {title} {Geometry of reduced density matrices for symmetry-protected
  topological phases},\ }\href {https://doi.org/10.1103/PhysRevA.93.012309}
  {\bibfield  {journal} {\bibinfo  {journal} {Physical Review A}\ }\textbf
  {\bibinfo {volume} {93}},\ \bibinfo {pages} {012309} (\bibinfo {year}
  {2016})}\BibitemShut {NoStop}%
\bibitem [{\citenamefont {Warren}\ and\ \citenamefont {Mazziotti}(2022)}]{dc}%
  \BibitemOpen
  \bibfield  {author} {\bibinfo {author} {\bibfnamefont {S.}~\bibnamefont
  {Warren}}\ and\ \bibinfo {author} {\bibfnamefont {D.~A.}\ \bibnamefont
  {Mazziotti}},\ }\href@noop {} {\bibinfo {title}
  {{2-RDM-Sets-from-Dual-Condensate-Hamiltonian}}},\ \bibinfo {howpublished}
  {\url{https://github.com/damazz/2-RDM-Sets-from-Dual-Condensate-Hamiltonian}}
  (\bibinfo {year} {2022}),\ \bibinfo {note} {[Online; accessed
  13-August-2022]}\BibitemShut {NoStop}%
\bibitem [{\citenamefont {Sager}\ and\ \citenamefont
  {Mazziotti}(2022{\natexlab{c}})}]{sager_entangled_2022}%
  \BibitemOpen
  \bibfield  {author} {\bibinfo {author} {\bibfnamefont {L.~M.}\ \bibnamefont
  {Sager}}\ and\ \bibinfo {author} {\bibfnamefont {D.~A.}\ \bibnamefont
  {Mazziotti}},\ }\bibfield  {title} {\bibinfo {title} {Entangled phase of
  simultaneous fermion and exciton condensations realized},\ }\href
  {https://doi.org/10.1103/PhysRevB.105.L121105} {\bibfield  {journal}
  {\bibinfo  {journal} {Physical Review B}\ }\textbf {\bibinfo {volume}
  {105}},\ \bibinfo {pages} {L121105} (\bibinfo {year}
  {2022}{\natexlab{c}})}\BibitemShut {NoStop}%
\bibitem [{Note1()}]{Note1}%
  \BibitemOpen
  \bibinfo {note} {See Supplemental Material at [URL will be inserted by
  publisher] for ground-state energies and their derivatives along the quantum
  phase transitions.}\BibitemShut {Stop}%
\bibitem [{\citenamefont {Bouscher}\ \emph {et~al.}(2022)\citenamefont
  {Bouscher}, \citenamefont {Panna}, \citenamefont {Balasubramanian},
  \citenamefont {Cohen}, \citenamefont {Ritter},\ and\ \citenamefont
  {Hayat}}]{bouscher_enhanced_2022}%
  \BibitemOpen
  \bibfield  {author} {\bibinfo {author} {\bibfnamefont {S.}~\bibnamefont
  {Bouscher}}, \bibinfo {author} {\bibfnamefont {D.}~\bibnamefont {Panna}},
  \bibinfo {author} {\bibfnamefont {K.}~\bibnamefont {Balasubramanian}},
  \bibinfo {author} {\bibfnamefont {S.}~\bibnamefont {Cohen}}, \bibinfo
  {author} {\bibfnamefont {D.}~\bibnamefont {Ritter}},\ and\ \bibinfo {author}
  {\bibfnamefont {A.}~\bibnamefont {Hayat}},\ }\bibfield  {title} {\bibinfo
  {title} {Enhanced {Cooper}-{Pair} {Injection} into a {Semiconductor}
  {Structure} by {Resonant} {Tunneling}},\ }\href
  {https://doi.org/10.1103/PhysRevLett.128.127701} {\bibfield  {journal}
  {\bibinfo  {journal} {Physical Review Letters}\ }\textbf {\bibinfo {volume}
  {128}},\ \bibinfo {pages} {127701} (\bibinfo {year} {2022})},\ \bibinfo
  {note} {publisher: American Physical Society}\BibitemShut {NoStop}%
\end{thebibliography}%
\end{document}